\documentclass[journal]{IEEEtran}
\newcommand{\leftscript}[3]{{\vphantom{#3}}^{#1}_{#2}{\!#3}}

\newcommand{\leftscriptspace}[3]{{\vphantom{#3}}^{#1}_{#2}{#3}}

\usepackage{cite}

\ifCLASSINFOpdf
\else
\fi
\usepackage[pdftex]{graphicx}
\usepackage[cmex10]{amsmath}
\usepackage{amssymb}
\usepackage{array}
\usepackage{color}
\usepackage{threeparttable}

\begin{document}
%
\title{Passive Lossless Huygens Metasurfaces for Conversion of Arbitrary Source Field to Directive Radiation}
%
%
%

\author{Ariel~Epstein,~\IEEEmembership{Member,~IEEE,}
        and~George~V.~Eleftheriades,~\IEEEmembership{Fellow,~IEEE}
\thanks{The authors are with the Edward S. Rogers Sr. Department of Electrical and
Computer Engineering, University of Toronto, Toronto, ON, Canada M5S 2E4 (e-mail: ariel.epstein@utoronto.ca; gelefth@waves.utoronto.ca).}
\thanks{Manuscript received March 27, 2014; revised August 27, 2014.}}

%
%

\markboth{IEEE Transactions on Antennas and Propagation,~Vol.~62, No.~11, November~2014}%
{Epstein and Eleftheriades: Passive Lossless Huygens Metasurfaces for Directive Radiation}
%

\IEEEpubid{0000--0000/00\$00.00~\copyright~2014 IEEE}


\maketitle

\begin{abstract}
We present a \textcolor{black}{semi-analytical} formulation of the interaction between a given source field and a scalar Huygens metasurface (HMS), a recently introduced promising concept for wavefront manipulation based on a sheet of orthogonal electric and magnetic dipoles. Utilizing the equivalent surface impedance representation of these metasurfaces, we establish that an arbitrary source field can be converted into directive radiation via a passive lossless HMS if two physical conditions are met: local power conservation and local impedance \textcolor{black}{equalization}. Expressing the fields via their plane-wave spectrum and harnessing the slowly-varying \textcolor{black}{envelope} approximation we obtain semi-analytical formulae for the scattered fields, and prescribe the surface reactance required for the metasurface implementation. The resultant design procedure indicates that the local impedance \textcolor{black}{equalization} induces a Fresnel-like reflection, while local power conservation forms a radiating virtual aperture which follows the total excitation field magnitude. The semi-analytical predictions are verified by finite-element simulations of HMSs designed for different source configurations. Besides serving as a flexible design procedure for HMS radiators, the proposed formulation also provides a \textcolor{black}{robust} mechanism to incorporate a variety of source configurations into general HMS models, as well as physical insight on the conditions enabling purely reactive implementation of this novel type of metasurfaces.
\end{abstract}


%
\IEEEpeerreviewmaketitle

\section{Introduction}
\label{sec:introduction}
%
%
%
%
\IEEEPARstart{E}{lectrically} thin sheets with repetitive metallic inclusions or exclusions have been used extensively in the past in antenna applications to control the properties of reflected or transmitted power, e.g. its direction, phase, or polarization \cite{Trentini1956, Pozar1997, Sievenpiper1999, Romeu2000, Sarabandi2007, Hum2014}. Such surfaces have received increasing attention lately, as part of the intensive research in the field of optical and microwave metamaterials, in an attempt to harness ideas from bulk metamterial explorations to design low-profile components with extraordinary wavefront manipulation capabilities \cite{Maci2011, Holloway2012, Kildishev2013, Yu2014}. In contrast to bulk metamaterials, where subwavelength elements are combined to form a volumetric entity with prescribed local response to electromagnetic fields, in metamaterial sheets, or metasurfaces, these subwavelength atomic units are confined to a region with subwavelength thickness. This geometrical difference should decrease significantly fabrication complexity of metasurfaces and also loss-related problems; however, it requires development of new design methodologies, as the interaction of electromagnetic fields with metasurfaces is naturally described via effective boundary conditions \cite{Kuester2003, Tretyakov2003, Zhao2011}, as opposed to effective permeabilities and permittivities (or effective wave equations), more suitable for modelling volumetric metamaterials \cite{Pendry2012, Eleftheriades2012, Martini2014}.

In particular, it was recently recognized that as metasurfaces act as sources of tangential field discontinuities, they can be modelled by a distribution of electric and magnetic surface currents, prescribed by the equivalence principle [\citenum{Balanis1997}, pp. 575-579]. Hence, in principle, for a given incident field, a desirable electromagnetic field distribution in space can be achieved by engineering the surface to induce currents that would produce the required tangential fields on both of its facets.

Approximating the required continuous surface currents by a dense distribution of electric and magnetic dipoles, passive surfaces implementing plane-wave refraction were demonstrated \cite{Pfeiffer2013, Selvanayagam2013}. The elementary sources were formed by subwavelength inductive and capacitive elements that produced the suitable magnitudes of the current in response to the exciting incident plane-wave. Simultaneously, it was demonstrated that also active elementary sources may be utilized to introduce desirable field discontinuities, e.g., to implement a cloaking device based on the same equivalence principle \cite{Selvanayagam2012, Selvanayagam2013, Selvanayagam2013_1}. As these surfaces were composed of orthogonal electric and magnetic dipoles engineered to induce unidirectional radiation, i.e. acting as Huygens sources [\citenum{Balanis1997}, pp. 653-660],\cite{Jin2010}, they were named Huygens metasurfaces (HMS) \cite{Pfeiffer2013}. 

\IEEEpubidadjcol
In addition to plane-wave refraction and cloaking, recent reports proposed designs of Huygens metasurfaces which implement beam shaping, transmission or reflection coefficient engineering and polarization manipulation (using tensor Huygens metasurfaces) \cite{Pfeiffer2013, Niemi2013, Rapoport2013, Selvanayagam2014}. Although the design methodologies differ between the various authors, they all rely on the fact that if the dimensions of the unit cells and their spatial arrangement obey certain conditions, the metasurface can be modelled by effective electric and magnetic polarizability distributions, which are translated to position-dependent sheet boundary conditions \cite{Kuester2003,Tretyakov2003,Holloway2005,Holloway2012}. These, in turn, can be equivalently described as surface impedance and surface admittance matrices relating the electric and magnetic field components at the two facets of the metasurface \cite{Pfeiffer2013,Selvanayagam2013}.


Following this approach greatly simplifies device design, as it facilitates the development of simple circuit models to Huygens metasurfaces \cite{Selvanayagam2013_1}. Moreover, as the effective surface impedance and admittance matrices are directly related to the \emph{locally averaged} polarizabilities of the elementary scatterers, we may assess the local equivalent surface impedance of a unit cell by simulating or measuring the effective impedance of an infinite periodic array of such identical unit cells
(local periodicity approximation) \cite{Tretyakov2003,Kuester2003,Holloway2005,Fong2010,Maci2011,Pfeiffer2013_1}.

Indeed, this modelling approach was used in recent demonstrations of HMSs: given the incident field and the desirable transmitted (or reflected) field, the required boundary conditions to support the field discontinuities can be formulated, resulting in the required surface impedance and admittance matrices (or electric and magnetic polarizability distributions).  

However, for realizing the Huygens metasurfaces, it is desirable for the elements implementing the required polarizability distributions to be passive and lossless, i.e. the surface impedance and admittance to possess pure imaginary values. Implementing impedance or admittance sheets with nonvanishing real parts requires engineering of gain or loss elements, thereby complicating greatly the design and realization. Nonetheless, following the simplistic methodology in which the discontinuities between the desirable and incident fields are directly translated into surface impedances and admittances by no means guarantees the passivity of the resultant metasurface (See, e.g., \cite{Pfeiffer2013,Selvanayagam2013}). In fact, in \cite{Pfeiffer2013} the design procedure did not consider the passivity limitation, however the resultant \textcolor{black}{complex surface impedance and admittance were such that they could be approximated by purely imaginary functions,}
leading to a well-functioning, \textcolor{black}{passive and lossless,} prototype. 
The reasons for that encouraging outcome were not analyzed therein, though.  

In addition, almost all Huygens metasurfaces presented in the literature to date were designed to be excited by a plane-wave or a beam propagating towards the surface in homogeneous medium \cite{Holloway2005, Pfeiffer2013, Selvanayagam2013, Niemi2013}. Nevertheless, to facilitate the development of realizable antenna devices based on Huygens metasurfaces it would be necessary to extend the current design techniques to enable excitation of the metasurface by localized (impulsive) sources, or waveguided modes. A step in that direction was made by Holloway \textit{et al.} \cite{Holloway2012_1}, where a line source excitation was considered, but only for metasurfaces with \emph{constant} polarizability density;
as demonstrated by \cite{Pfeiffer2013, Selvanayagam2013}, allowing the polarizability densities to vary along the metasurface could be beneficial, providing more degrees of freedom for the design.

In this work, we derive \textcolor{black}{from first principles} simple rules for designing passive lossless Huygens metasurfaces producing directive radiation to a prescribed angle when excited by a given (arbitrary) source field. Decomposing the fields to their plane-wave spectrum and generalizing the approach presented in \cite{Selvanayagam2013} for plane-wave excitation, we show that satisfying two physical conditions is sufficient to guarantee that the desirable functionality can be achieved by purely reactive surfaces: local power conservation across the surface, and local impedance \textcolor{black}{equalization of} the fields on both sides of the metasurface. Enforcing these conditions locally, i.e. at each point on the surface, leads to a complementary set of simple expressions for the surface impedance and surface admittance, as well as facilitates the semi-analytical evaluation of the reflected and transmitted fields. These results enable design of directive HMS radiators with a wide range of source excitations, thus extending significantly the possible applications. Moreover, the derivation provides clear physical interpretation of the conditions required to implement passive lossless HMSs, which may also be indicative to equivalent requirements in more generalized scenarios (e.g. HMSs which perform other functionalities). Altogether, this forms an efficient and powerful tool for HMS engineering, promoting design of novel antenna devices.               

\section{Theory}
\label{sec:theory}
\subsection{Formulation}
\label{subsec:theory:formulation}
\begin{figure}[!t]
\centering
\includegraphics[width=8.5cm]{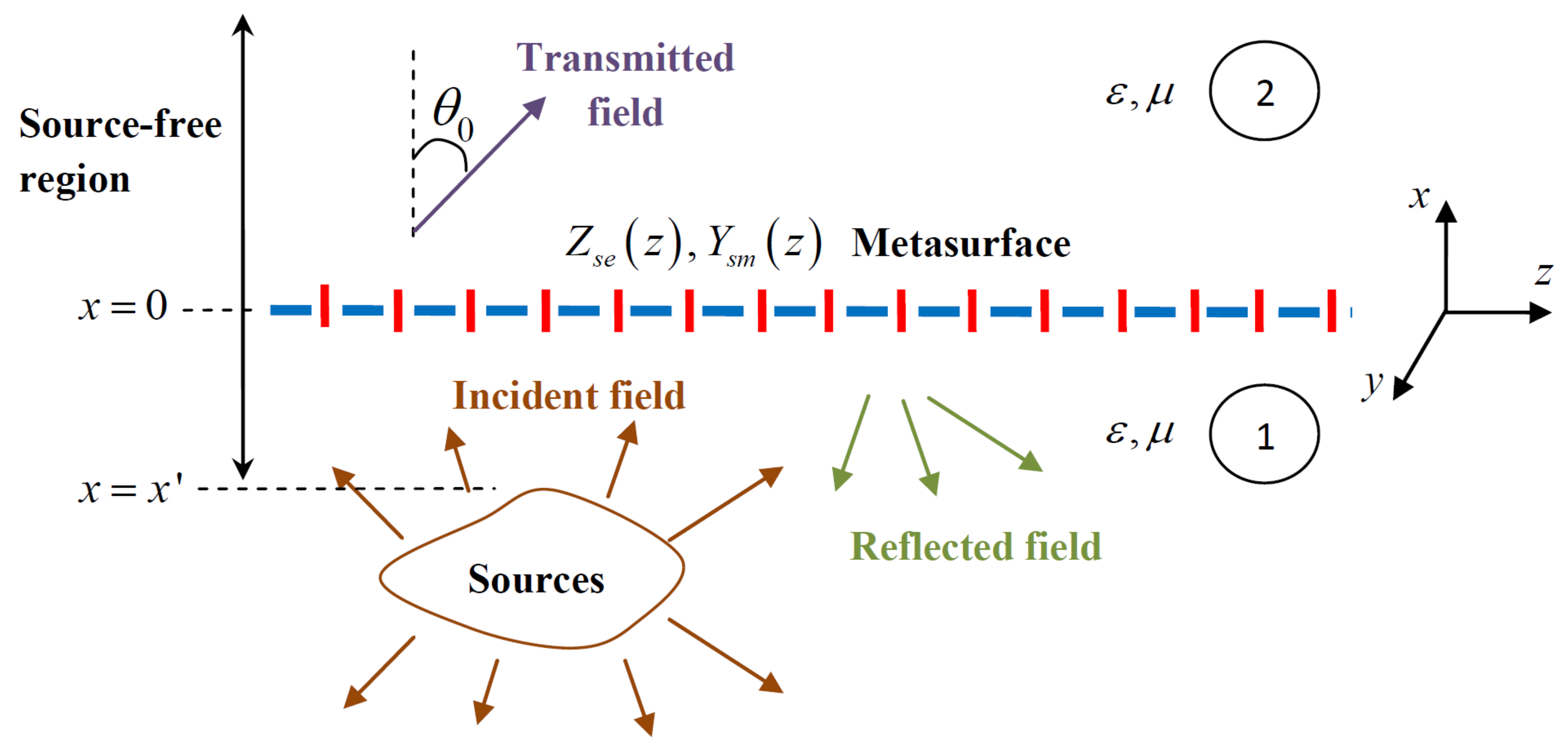}
\caption{Physical configuration of a Huygens metasurface excited by an arbitrary source situated at $x\leq x'<0$. The formalism applies without modification also to scenarios in which the region $x\leq x'$ is occupied by an inhomogeneous medium, as long as its cross-section remains uniform with respect to the $x$ axis, i.e the permittivity, permeability and conductivity are a function of $x$ coordinate only (e.g., plane-stratified media) [\citenum{FelsenMarcuvitz1973}, pp. 183-202].}
\label{fig:PhysicalConfiguration}
\end{figure}

We consider a 2D configuration $\partial/\partial y =0$ in which a Huygens metasurface situated at $x=0$ is excited by an arbitrary current distribution limited to the half-space $x<0$ (Fig. \ref{fig:PhysicalConfiguration}). The surrounding media is assumed to be homogeneous, with permittivity $\epsilon$ and permeability $\mu$, defining the wave impedance $\eta=\sqrt{\mu/\epsilon}$. Harmonic time dependency of $e^{j\omega t}$ is assumed (and suppressed), defining the wavenumber $k=\omega\sqrt{\epsilon\mu}$.

The \textcolor{black}{metasurface} is characterized by its surface impedance $Z_{se}\left(z\right)$ and surface admittance $Y_{sm}\left(z\right)$, inducing discontinuities in tangential magnetic and electric field components, respectively, given by the generalized sheet transition conditions (GSTC) as formulated by Kuester \textit{et al.} \cite{Kuester2003}
\begin{eqnarray}
\left\lbrace\!\!\!
\begin{array}{l}
Z_{se}\left(z\right)\vec{J}_s = 
	Z_{se}\left(z\right)\hat{x}\times\left[\left.\vec{H}\right|_{x\rightarrow0^{+}}-\left.\vec{H}\right|	_{x\rightarrow0^{-}}\right] =  \\
\bigskip =\dfrac{1}{2}\left[\left.\vec{E}\right|_{x\rightarrow0^{+}}+\left.\vec{E}\right|_{x\rightarrow0^{-}}\right]
 \\
Y_{sm}\left(z\right)\vec{M}_s = 
	-Y_{sm}\left(z\right)\hat{x}\times\left[\left.\vec{E}\right|_{x\rightarrow0^{+}}-\left.\vec{E}\right|_{x\rightarrow0^{-}}\right] =  \\
\bigskip = \dfrac{1}{2}\left[\left.\vec{H}\right|_{x\rightarrow0^{+}}+\left.\vec{H}\right|_{x\rightarrow0^{-}}\right],
\end{array}
\right.\!\!\!\!\!\!\!
\label{equ:general_impedance_admittance_discontinuity}
\end{eqnarray}
where $\vec{J}_s$ and $\vec{M}_s$ are the surface currents induced by the tangential electric and magnetic field components, respectively, and we assumed the impedance and admittance matrices can be described by scalar quantities. The half-spaces below and above the metasurface are referred to as region 1 and region 2, respectively, and we require the sources not to be infinitesimally close to the HMS such that a source-free region $x'<x<0$ can be defined just below it (Fig. \ref{fig:PhysicalConfiguration}). 

\textcolor{black}{To ensure that we harness all possible degrees of freedom for the HMS design, we wish to analyze the most general field constellation admissible by Maxwell's equations in our 2D scenario.
This is achieved by considering both TE and TM polarized fields, by allowing reflections where applicable, and by utilizing the spectral representation of the fields for semi-analytical formulation. To facilitate the fluent reading of the rest of the paper, we include the basic field definitions in this Subsection. Starting with Maxwell's equations, these definitions enable us to precisely point out the approximations made along the way, guaranteeing the consistency of the derivation and the accurate interpretation of its results.}

In region 1, \textcolor{black}{thus,} we distinguish between the incident field and reflected field, while region 2 is populated only by the transmitted fields (denoted by $\mathrm{inc}$, $\mathrm{ref}$, and $\mathrm{trans}$ superscripts, respectively)
\begin{eqnarray}
&\vec{E}\left(x,z\right)=
\left\lbrace
\begin{array}{l l}
\vec{E}^{\mathrm{inc}}\left(x,z\right)+\vec{E}^{\mathrm{ref}}\left(x,z\right) & x<0 \\
\vec{E}^{\mathrm{trans}}\left(x,z\right) & x>0 
\end{array}
\right. \nonumber \\ 
&\vec{H}\left(x,z\right)=
\left\lbrace
\begin{array}{l l}
\vec{H}^{\mathrm{inc}}\left(x,z\right)+\vec{H}^{\mathrm{ref}}\left(x,z\right) & x<0 \\
\vec{H}^{\mathrm{trans}}\left(x,z\right) & x>0, 
\end{array}
\right.
\label{equ:general_incident_reflected_transmitted}
\end{eqnarray}
and the incident fields are defined as the fields produced by the sources in the absence of the HMS\footnote{A formal definition of the incident fields, more suitable in cases the region $x<x'$  includes scattering elements (e.g., as we allow in Subsection \ref{subsec:results:line_source_PEC}) will be given in \eqref{equ:TE_spectral_domain_Ey}-\eqref{equ:TE_spectral_domain_Hz}.}.

In 2D configurations Maxwell's equations can be separated to two decoupled sets of equations related to the transverse electric (TE) or transverse magnetic (TM) polarized field components. In source-free regions, the TE-polarized ($E_x=0$) nonvanishing field components are $\left(E_y,H_x,H_z\right)$, and Maxwell's equations can be reduced to a wave equation for $E_y$ and curl equations relating the other components to it,
\begin{eqnarray}
\!\!\begin{array}{l l l}
\left(\nabla^2+k^2\right)E_y=0, & H_x=\frac{1}{jk\eta}\frac{\partial E_y}{\partial z}, & 
H_z=-\frac{1}{jk\eta}\frac{\partial E_y}{\partial x}.
\end{array}\!\!\!
\label{equ:TE_Maxwell_equations}
\end{eqnarray}
Analogously, in source-free regions, the TM-polarized ($H_x=0$) nonvanishing field components are $\left(H_y,E_x,E_z\right)$, and Maxwell's equations can be reduced to 
\begin{eqnarray}
\!\!\begin{array}{l l l}
\left(\nabla^2+k^2\right)H_y=0, & E_x=-\frac{\eta}{jk}\frac{\partial H_y}{\partial z},
 & E_z=\frac{\eta}{jk}\frac{\partial H_y}{\partial x}.
\end{array}\!\!\!
\label{equ:TM_Maxwell_equations}
\end{eqnarray}

As this work does not deal with polarization manipulation of the source fields, we may design an HMS directive radiator for each polarization independently, namely a TE-HMS and a TM-HMS. The TE-HMS would interact only with $E_y$ and $H_z$, therefore the required surface impedance $Z_{se}\left(z\right)$ would be implemented using scatterers sensitive to electric field in the $y$ direction (e.g. loaded wires parallel to the $y$-axis) and the surface admittance $Y_{sm}\left(z\right)$  would be implemented using scatterers sensitive to magnetic field in the $z$ direction (e.g. loaded loops whose axis is parallel to the $z$-axis) \cite{Selvanayagam2013}. On the other hand, the TM-HMS should interact only with $H_y$ and $E_z$, therefore $Z_{se}\left(z\right)$ and $Y_{sm}\left(z\right)$ would be composed by scatterers sensitive to electric field in the $z$ direction and magnetic field in the $y$ direction, respectively. As the elements implementing the TE-HMS and the TM-HMS are orthogonal, there should be no coupling between them; thus, ideally, the two HMSs may be combined to a single metasurface without any changes to the independent designs. In view of this observation, we formulate the procedure to design an HMS assuming the incident field is TE-polarized; the design rules for TM-polarized excitation can be readily derived by duality (Appendix \ref{app:theory_TM_excitation}). Extension of this work to polarization manipulating HMSs \cite{Selvanayagam2014} will be addressed in a separate report.  
%

\subsubsection{Spectral Decomposition}
\label{subsubsec:theory:formulation:spectral_decomposition}
As denoted, when the excitation field is TE-polarized, the nonvanishing field components are $E_y\left(x,z\right)$, $H_z\left(x,z\right)$, and $H_x\left(x,z\right)$; the scalar surface impedance only induces \emph{electric} currents in the $y$ direction; and the scalar surface admittance only induces \emph{magnetic} currents in the $z$ direction. In view of \eqref{equ:TE_Maxwell_equations} a general solution for the fields in the \emph{source-free region} $x>x'$ can be formulated in the spectral domain as \cite{FelsenMarcuvitz1973, Chew1990, Holloway2012_1, Epstein2013_3}
\begin{eqnarray}
\left\lbrace
\begin{array}{l}
\!\!\!\!E_y^{\mathrm{inc}}\left(x,z\right) = k\eta \dfrac{I_0}{2\pi}\displaystyle\int\limits_{-\infty}^{\infty}\dfrac{dk_t}{2\beta}
	\leftscript{e}{}{f}\left(k_t\right)e^{-j\beta x}e^{jk_tz} \\
\!\!\!\!E_y^{\mathrm{ref}}\left(x,z\right) = -k\eta \dfrac{I_0}{2\pi}\displaystyle\int\limits_{-\infty}^{\infty}\dfrac{dk_t}{2\beta}
	\leftscriptspace{e}{}{\Gamma}\left(k_t\right)\leftscript{e}{}{f}\left(k_t\right)e^{j\beta x}e^{jk_tz} \\
\!\!\!\!E_y^{\mathrm{trans}}\left(x,z\right) = k\eta \dfrac{I_0}{2\pi}\displaystyle\int\limits_{-\infty}^{\infty}\dfrac{dk_t}{2\beta}
	\leftscriptspace{e}{}{T}\left(k_t\right)e^{-j\beta x}e^{jk_tz},
\end{array}\!\!\!\!\!
\right.
\label{equ:TE_spectral_domain_Ey}
\end{eqnarray}
where $k_t$ is the transverse wavenumber (associated with the propagation along $z$), and $\beta=\sqrt{k^2-k_t^2}$ is the longitudinal wavenumber (associated with the propagation along $x$); to satisfy the radiation condition we demand that $\Im\left\lbrace\beta\right\rbrace<0$. The $e$ left superscript denotes TE-HMS related parameters throughout the paper, and $I_0$ is a unit current magnitude. 

It can be readily verified that all three integrals in \eqref{equ:TE_spectral_domain_Ey} satisfy the wave equation of \eqref{equ:TE_Maxwell_equations}, where $\leftscript{e}{}{f}\left(k_t\right)$ is the source-related plane-wave spectrum, $\leftscriptspace{e}{}{\Gamma}\left(k_t\right)$ is the reflection coefficient in the spectral domain, and $\leftscriptspace{e}{}{T}\left(k_t\right)$ corresponds to the spectral content of the transmitted fields. 
As the sources reside at $x\leq x'$ in region 1, and the scattering metasurface is situated at $x=0$, the general solution in $x'<x<0$ must consider both upwards (incident) and downwards (reflected) propagating waves. In contrast, as neither sources nor scatterers exist in region 2, only upwards (transmitted) propagating waves are allowed for $x>0$, as to satisfy the radiation condition.

We emphasize that the formal solution presented in \eqref{equ:TE_spectral_domain_Ey} is valid in general only for $x>x'$ and that the presence of the sources at $x\leq x'$ introduces a discontinuity in the fields, which should be accounted for in that region. 
Nonetheless, for designing the HMS only the tangential fields at $x\rightarrow 0^{\pm}$ as formulated in \eqref{equ:TE_spectral_domain_Ey} are required \cite{Holloway2012_1}, and this formulation is valid \emph{regardless} of the nature of the source distribution $x\leq x'$. In fact,  \eqref{equ:TE_spectral_domain_Ey} applies without modification also to scenarios in which the region $x\leq x'$ is occupied by an inhomogeneous medium, as long as its cross-section remains uniform with respect to the $x$ axis, i.e the permittivity, permeability and conductivity are a function of $x$ coordinate only (e.g., plane-stratified media) [\citenum{FelsenMarcuvitz1973}, pp. 183-202]. As shall be demonstrated in Subsection \ref{subsec:results:line_source_PEC}, this allows for even a wider range of excitation schemes to be investigated using our model. 

From \eqref{equ:TE_spectral_domain_Ey} and the curl equations of \eqref{equ:TE_Maxwell_equations} we derive the corresponding tangential component of the magnetic field at $x>x'$,
\begin{eqnarray}
\left\lbrace
\begin{array}{l}
\!\!\!\!H_z^{\mathrm{inc}}\left(x,z\right) = \dfrac{I_0}{2\pi}\displaystyle\int\limits_{-\infty}^{\infty}\dfrac{dk_t}{2}
	\leftscript{e}{}{f}\left(k_t\right)e^{-j\beta x}e^{jk_tz} \\
\!\!\!\!H_z^{\mathrm{ref}}\left(x,z\right) = \dfrac{I_0}{2\pi}\displaystyle\int\limits_{-\infty}^{\infty}\dfrac{dk_t}{2}
	\leftscriptspace{e}{}{\Gamma}\left(k_t\right)\leftscript{e}{}{f}\left(k_t\right)e^{j\beta x}e^{jk_tz} \\
\!\!\!\!H_z^{\mathrm{trans}}\left(x,z\right) = \dfrac{I_0}{2\pi}\displaystyle\int\limits_{-\infty}^{\infty}\dfrac{dk_t}{2}
	\leftscriptspace{e}{}{T}\left(k_t\right)e^{-j\beta x}e^{jk_tz},
\end{array}\!\!\!\!
\right.
\label{equ:TE_spectral_domain_Hz}
\end{eqnarray}
and \eqref{equ:general_impedance_admittance_discontinuity} can be rewritten as \cite{Selvanayagam2013}
\begin{eqnarray}
\!\left\lbrace
\begin{array}{l}
\vspace{5pt}
\!\!\!\!\leftscript{e}{}{Z}_{se}\left(z\right)=-\dfrac{1}{2}
\dfrac{E_y^{\mathrm{trans}}\left(0,z\right)+\left[E_y^{\mathrm{inc}}\left(0,z\right)+E_y^{\mathrm{ref}}\left(0,z\right)\right]}{H_z^{\mathrm{trans}}\left(0,z\right)-\left[H_z^{\mathrm{inc}}\left(0,z\right)+H_z^{\mathrm{ref}}\left(0,z\right)\right]} \\
\!\!\!\!\!\leftscriptspace{e}{}{Y}_{sm}\left(z\right)=-\dfrac{1}{2}
\dfrac{H_z^{\mathrm{trans}}\left(0,z\right)+\left[H_z^{\mathrm{inc}}\left(0,z\right)+H_z^{\mathrm{ref}}\left(0,z\right)\right]}{E_y^{\mathrm{trans}}\left(0,z\right)-\left[E_y^{\mathrm{inc}}\left(0,z\right)+E_y^{\mathrm{ref}}\left(0,z\right)\right]}
\end{array}\!\!\!\!\!\!\!\!\!
\right.
\label{equ:TE_impedance_admittance_discontinuity}
\end{eqnarray}

\subsubsection{Statement of the Problem}
\label{subsubsec:theory:formulation:problem_statement}
Given an incident field, we would like to determine the required variation of the surface impedance and admittance along the metasurface such that
\begin{itemize}
	\item Both the surface impedance and surface admittance are purely reactive (passive and lossless), namely ${\Re\left\lbrace Z_{se}\left(z\right)\right\rbrace=\Re\left\lbrace Y_{sm}\left(z\right)\right\rbrace=0}$. For that we are willing to allow some reflections from the metasurface.
	\item The transmitted field $E_y^{\mathrm{trans}}\left(x,z\right)$, $H_z^{\mathrm{trans}}\left(x,z\right)$, $H_x^{\mathrm{trans}}\left(x,z\right)$ will form a directional radiation to a specified angle $\theta_0$ with respect to the $x$ axis (Fig. \ref{fig:PhysicalConfiguration}). To this end we would like to form a virtual aperture on the surface with surface current having the suitable linear phase variation and as uniform as possible magnitude.
\end{itemize}

In order to satisfy the second demand, we require that the spectral content of the transmitted fields $\leftscriptspace{e}{}{T}\left(k_t\right)$ will be localized around $k_t=k_{t,0}=-k\sin\theta_0$.
\textcolor{black}{Ideally, $\leftscriptspace{e}{}{T}\left(k_t\right)$ would be a delta function; 
however, this can be obtained only for uniform excitation of an infinite HMS, i.e. for plane waves (e.g., as in \cite{Selvanayagam2013}). When finite sources and metasurfaces are considered, the virtual aperture must have a compact support in space. To facilitate this, we introduce to our formulation a slowly-varying window function $\leftscriptspace{e}{}{W\left(x,z\right)}$, decaying towards the edges of the metasurface, which would serve as an envelope for the desirable linear phase function. If the envelope variation would be moderate with respect to the required phase variation, the resulting radiation would be directed as desired. Formally, we define this virtual aperture window function as}
\begin{eqnarray}
E_y^{\mathrm{trans}}\left(x,z\right) \triangleq k\eta I_0 \leftscriptspace{e}{}{W\left(x,z\right)}e^{-jkx\cos\theta_0}e^{-jkz\sin\theta_0}
\label{equ:TE_window_function_Ey}
\end{eqnarray}
and demand that the HMS would be designed such that the resulting virtual aperture window would form a slowly-varying \textcolor{black}{envelope} as $x\rightarrow 0^{+}$, namely,
\begin{eqnarray}
\left|\frac{\partial}{\partial x}\leftscriptspace{e}{}{W\left(x,z\right)}\right|_{x\rightarrow 0^+}
	\ll\left|k\cos\theta_0\leftscriptspace{e}{}{W\left(x,z\right)}\right|_{x\rightarrow 0^+}.
\label{equ:TE_slowly_varying_W}
\end{eqnarray}
Applying this constraint on \eqref{equ:TE_spectral_domain_Hz} enables us to approximate the tangential magnetic field at $x\rightarrow 0^{+}$ as
\begin{eqnarray}
H_z^{\mathrm{trans}}\!\left(x,z\right)\! \approx \!k\cos\theta_0I_0\leftscriptspace{e}{}{W\left(x,z\right)}e^{-jkx\cos\theta_0}e^{-jkz\sin\theta_0}\!.\!
\label{equ:TE_window_function_Hz}
\end{eqnarray}
\textcolor{black}{The condition \eqref{equ:TE_slowly_varying_W}, thus, ensures the tangential fields on the virtual aperture locally resemble those of a plane-wave towards $\theta_0$, promoting directive radiation.}

Substituting \eqref{equ:TE_window_function_Ey} and \eqref{equ:TE_window_function_Hz} into \eqref{equ:TE_impedance_admittance_discontinuity} yields
\begin{eqnarray}
\left\lbrace
\begin{array}{l}
\vspace{5pt}
\!\!\!\!\leftscript{e}{}{Z}_{se}\left(z\right)=-\dfrac{\eta}{2\cos\theta_0}
\dfrac{\leftscript{e}{}{F}^{+}\left(z\right)+\leftscript{e}{}{F}_{\!\!E}^{-}\left(z\right)}
{\leftscript{e}{}{F}^{+}\left(z\right)-\leftscript{e}{}{F}_{\!\!H}^{-}\left(z\right)} \\
\!\!\!\!\leftscriptspace{e}{}{Y}_{sm}\left(z\right)=-\dfrac{\cos\theta_0}{2\eta}
\dfrac{\leftscript{e}{}{F}^{+}\left(z\right)+\leftscript{e}{}{F}_{\!\!H}^{-}\left(z\right)}
{\leftscript{e}{}{F}^{+}\left(z\right)-\leftscript{e}{}{F}_{\!\!E}^{-}\left(z\right)},
\end{array}\!\!\!\!\!
\right.
\label{equ:TE_impedance_admittance_discontinuity_dimensionless}
\end{eqnarray}
where we have defined dimensionless quantities, proportional to the fields on the lower (minus-sign superscript) and upper (plus-sign superscript) facets of the metasurface, as follows
\begin{eqnarray}
\left\lbrace
\begin{array}{l}
\vspace{5pt}
\leftscript{e}{}{F}_{\!\!E}^{-}\left(z\right) \triangleq \dfrac{1}{I_0k\eta}
	\left[E_y^{\mathrm{inc}}\left(0,z\right)+E_y^{\mathrm{ref}}\left(0,z\right)\right] \\
\vspace{5pt}
\leftscript{e}{}{F}_{\!\!H}^{-}\left(z\right) \triangleq \dfrac{1}{I_0k\cos\theta_0}
	\left[H_z^{\mathrm{inc}}\left(0,z\right)+H_z^{\mathrm{ref}}\left(0,z\right)\right] \\
\leftscript{e}{}{F}^{+}\left(z\right) \triangleq \leftscriptspace{e}{}{W\left(0,z\right)}e^{-jkz\sin\theta_0}.
\end{array}\!\!\!\!\!\!
\right.
\label{equ:TE_dimensionless_quantities}
\end{eqnarray}

\textcolor{black}{It should be noted that the fields in our formulation are "macroscopic" in the sense that they result from averaged boundary conditions, applicable for an \emph{infinitely-thin} \emph{homogenized} \emph{equivalent} surface, exhibiting \emph{continuous} sheet impedance and admittance profiles \cite{Holloway2012}. Hence, they are strictly valid only from a finite distance away from the metasurface, where the subwavelength field variations due to the elements implementing the HMS become negligible. In general, this distance should be larger than both the unit cell thickness and periodicity, to ensure sufficient decay of corresponding higher-order Floquet modes [\citenum{Tretyakov2003}, pp.79-82].}

\textcolor{black}{
This is important, for instance, when interpreting the slowly-varying envelope condition \eqref{equ:TE_slowly_varying_W}, as the limit $x\rightarrow 0^{+}$ is only applicable for the "macroscopic" fields; nonetheless, as guaranteed by the GSTC derivation \cite{Tretyakov2003,Kuester2003}, adhering to this constraint when designing the HMS, would result in the desirable virtual aperture formation at the regions where the equivalent and real physical problems coincide \cite{Holloway2005,Holloway2012_1,Pfeiffer2013,Selvanayagam2013_1}. Practically, for the HMS implementation we utilize herein (Appendix \ref{app:HFSS_implementation}), keeping the sources and the observation points at least $\lambda/10$ away from the HMS plane $x=0$, should maintain the model accuracy.}

\subsection{Sufficient Conditions for Passive Lossless HMS}
\label{subsec:theory:passivity_sufficient_conditions}
\subsubsection{Local Power Conservation}
\label{subsubsec:theory:passivity_sufficient_conditions:power_conservation}
As we require the HMS to be purely reactive, the real power across the metasurface must be conserved. However, to obtain sufficient conditions for designing passive lossless HMSs we may require a stricter condition to be met, namely, that the real power impinging the metasurface from region 1 is equal \textit{locally} to the real power transmitted to region 2, at \emph{each point} $z$ on the surface.

To assess the consequences of this local power conservation condition we utilize \eqref{equ:TE_dimensionless_quantities} to express the local power densities along the metasurface in region 1 and 2 using the dimensionless field quantities. These are given, respectively, by the projection of the Poynting vector on the $x$ axis as $x\rightarrow 0^-$
\begin{eqnarray}
&\leftscript{e}{}{S}_x^-\left(z\right) = 
\frac{1}{2}\hat{x}\cdot\left[\vec{E}\left(x,z\right)\times\vec{H}^*\left(x,z\right)\right]_{x\rightarrow 0^-}
\nonumber \\
&\bigskip =\frac{1}{2}k\eta\left|I_0\right|^2\leftscript{e}{}{F}_{\!\!E}^{-}\left(z\right)
	\leftscript{e}{}{F}_{\!\!H}^{-*}\left(z\right)k\cos\theta_0,
\label{equ:TE_Poynting_vector_region1}
\end{eqnarray}
and as $x\rightarrow 0^+$
\begin{eqnarray}
&\leftscript{e}{}{S}_x^+\left(z\right) = 
\frac{1}{2}\hat{x}\cdot\left[\vec{E}\left(x,z\right)\times\vec{H}^*\left(x,z\right)\right]_{x\rightarrow 0^+}
\nonumber \\
&\bigskip =\frac{1}{2}k\eta\left|I_0\right|^2\left|\leftscript{e}{}{F}^{+}\left(z\right)\right|^2k\cos\theta_0,
\label{equ:TE_Poynting_vector_region2}
\end{eqnarray}
where the asterisk indicates the complex-conjugate operation.

The condition for local power conservation is $\Re\left\lbrace\leftscript{e}{}{S}_x^-\left(z\right)\right\rbrace=\Re\left\lbrace\leftscript{e}{}{S}_x^+\left(z\right)\right\rbrace$ which reads, using \eqref{equ:TE_Poynting_vector_region1}-\eqref{equ:TE_Poynting_vector_region2},
\begin{eqnarray}
\left|\leftscript{e}{}{F}^{+}\left(z\right)\right|^2=
	\Re\left\lbrace\leftscript{e}{}{F}_{\!\!E}^{-}\left(z\right)
	\leftscript{e}{}{F}_{\!\!H}^{-*}\left(z\right)\right\rbrace.
\label{equ:TE_local_power_conservation}
\end{eqnarray}
Multiplying the rational functions in \eqref{equ:TE_impedance_admittance_discontinuity_dimensionless} by the complex conjugate of their respective denominators and plugging in the local power conservation requirement \eqref{equ:TE_local_power_conservation} yields
\begin{eqnarray}
\left\lbrace
\begin{array}{l}
\vspace{5pt}
\!\!\!\!\leftscript{e}{}{Z}_{se}\left(z\right)\!=\!-\dfrac{\eta}{2\cos\theta_0}
\dfrac{\leftscript{e}{}{F}^{+*}\!\left(z\right)\leftscript{e}{}{F}_{\!\!E}^{-}\left(z\right)
			-\leftscript{e}{}{F}^{+}\left(z\right)\leftscript{e}{}{F}_{\!\!H}^{-*}\!\left(z\right)}
	{\left|\leftscript{e}{}{F}^{+}\left(z\right)-\leftscript{e}{}{F}_{\!\!H}^{-}\left(z\right)\right|^2} \\
\!\!\!\!\leftscriptspace{e}{}{Y}_{sm}\left(z\right)\!=\!-\dfrac{\cos\theta_0}{2\eta}
\dfrac{\leftscript{e}{}{F}^{+*}\!\left(z\right)\leftscript{e}{}{F}_{\!\!H}^{-}\left(z\right)
			-\leftscript{e}{}{F}^{+}\left(z\right)\leftscript{e}{}{F}_{\!\!E}^{-*}\!\left(z\right)}
	{\left|\leftscript{e}{}{F}^{+}\left(z\right)-\leftscript{e}{}{F}_{\!\!E}^{-}\left(z\right)\right|^2}
\end{array}\!\!\!\!\!\!\!\!\!\!
\right.
\label{equ:TE_impedance_admittance_discontinuity_power_conservation}
\end{eqnarray}

\subsubsection{Local Impedance \textcolor{black}{Equalization}}
\label{subsubsec:theory:passivity_sufficient_conditions:impedance_matching}
In view of \eqref{equ:TE_impedance_admittance_discontinuity_power_conservation}, a \emph{sufficient} condition for the real part of both $\leftscript{e}{}{Z}_{se}$ and $\leftscript{e}{}{Y}_{sm}$ to vanish is given by
\begin{eqnarray}
\leftscript{e}{}{F}_{\!\!E}^{-}\left(z\right)=
	\leftscript{e}{}{F}_{\!\!H}^{-}\left(z\right)\triangleq
	\leftscript{e}{}{F}^{-}\left(z\right),
\label{equ:TE_local_impedance_matching}
\end{eqnarray}
resulting in a purely imaginary numerator for both fractions.

The physical meaning of the latter is revealed by rewriting \eqref{equ:TE_local_impedance_matching} using \eqref{equ:TE_dimensionless_quantities}: the condition locally \textcolor{black}{equalizes} the wave impedance of the fields on the two facets of the metasurface. Indeed, substituting \eqref{equ:TE_dimensionless_quantities} into \eqref{equ:TE_local_impedance_matching} leads to the local impedance \textcolor{black}{equalization} condition
\begin{eqnarray}
\frac{E_y^{\mathrm{inc}}\left(0,z\right)+E_y^{\mathrm{ref}}\left(0,z\right)}{H_z^{\mathrm{inc}}\left(0,z\right)+H_z^{\mathrm{ref}}\left(0,z\right)}=
\frac{E_y^{\mathrm{trans}}\left(0,z\right)}{H_z^{\mathrm{trans}}\left(0,z\right)}=
	\frac{\eta}{\cos\theta_0},
\label{equ:TE_local_impedance_matching_explicit}
\end{eqnarray}
\textcolor{black}{and we stress that, for an arbitrary source, the existence of a reflected field is generally necessary to satisfy this condition.}

When \eqref{equ:TE_local_impedance_matching} is satisfied, the local power conservation condition \eqref{equ:TE_local_power_conservation} takes a simpler form, namely,
\begin{eqnarray}
\left|\leftscript{e}{}{F}^{+}\left(z\right)\right|^2=
\left|\leftscript{e}{}{F}^{-}\left(z\right)\right|^2\triangleq
\left|\leftscript{e}{}{F}\left(z\right)\right|^2.
\label{equ:TE_local_power_conservation_impedance_matched}
\end{eqnarray}   
Subsequently, we may define the dimensionless field quantities above and below the metasurface as
\begin{eqnarray}
\left\lbrace
\begin{array}{l}
	\leftscript{e}{}{F}^{+}\left(z\right) \triangleq \left|\leftscript{e}{}{F}\left(z\right)\right| 
		e^{j\varphi_{+}\left(z\right)} \\
	\leftscript{e}{}{F}^{-}\left(z\right) \triangleq \left|\leftscript{e}{}{F}\left(z\right)\right| 
		e^{j\varphi_{-}\left(z\right)},
\end{array}
\right.
\label{equ:TE_dimensionless_quantities_passive_HMS}
\end{eqnarray}
where $\varphi_{\pm}\left(z\right) \in \mathbb{R}$. Substituting these definitions into \eqref{equ:TE_impedance_admittance_discontinuity_power_conservation} leads to a complementary set of compact expressions for the desirable surface impedance and surface admittance
\begin{eqnarray}
\left\lbrace
\begin{array}{l}
\vspace{5pt}
\!\!\!\!\leftscript{e}{}{Z}_{se}\left(z\right)=-j\dfrac{\leftscript{e}{}{Z}_0}{2}
	\cot\left[\dfrac{\varphi_{-}\left(z\right)-\varphi_{+}\left(z\right)}{2}\right]
	 \\
\!\!\!\!\leftscriptspace{e}{}{Y}_{sm}\left(z\right)=-j\dfrac{\leftscriptspace{e}{}{Y}_0}{2}
	\cot\left[\dfrac{\varphi_{-}\left(z\right)-\varphi_{+}\left(z\right)}{2}\right],
\end{array}\!\!\!\!\!
\right.
\label{equ:TE_impedance_admittance_final}
\end{eqnarray}
where $\leftscript{e}{}{Z}_0=1/\leftscriptspace{e}{}{Y}_0=\eta/\cos\theta_0$ is the wave impedance of a TE-polarized plane-wave propagating in region 2 at an angle of $\theta_0$ with respect to the $x$ axis, given generally by [\citenum{FelsenMarcuvitz1973}]
\begin{eqnarray}
\vec{E}_t\left(0,z\right) \triangleq Z_0\left[\vec{H}_t\left(0,z\right)\times\hat{x}\right],
\label{equ:transmitted_wave_impedance}
\end{eqnarray}
$\vec{E}_t$ and $\vec{H}_t$ being the tangential components of the fields. As $\leftscript{e}{}{Z}_0$, $\leftscript{e}{}{Y}_0$, $\varphi_{\pm}\left(z\right)$ are all real, the HMS defined via \eqref{equ:TE_impedance_admittance_final} is indeed purely reactive, as required. 

\textcolor{black}{It is important to note that although it may appear that the required effect of the HMS is merely to introduce a local phase-shift to the incident field, seemingly allowing for a passive lossless implementation without any reflections, this is not the case, in general. In order to transform one valid electromagnetic field to another (both satisfying Maxwell's equations), one has to consider also the variation of the local wave impedance at each point along the metasurface. As demonstrated, for example, in \cite{Pfeiffer2013}, transforming both the local phase and the local wave impedance of the incident fields without incurring reflections in \eqref{equ:TE_impedance_admittance_discontinuity} gives rise to surface impedance and admittance values with non-vanishing real parts, which should be somehow mitigated if lossless passive realization is desirable (See "Supplemental Material" of \cite{Pfeiffer2013}, pp. 3-6). Nonetheless, as suggested in \cite{Selvanayagam2013} and generalized herein, introducing another degree of freedom in the form of the reflected fields to the design enables such a control, requiring neither active nor lossy elements.}

\textcolor{black}{\textcolor{black}{Lastly,} we emphasize that \emph{both} local power conservation \eqref{equ:TE_local_power_conservation} \emph{and} local impedance \textcolor{black}{equalization} \eqref{equ:TE_local_impedance_matching} are required to establish \emph{sufficient} conditions for a passive lossless HMS for our application. Moreover, our derivation results only in \emph{sufficient} conditions, thus it does not invalidate the possibility that other passive lossless designs, not adhering to these conditions, may achieve similar functionality.}

\subsection{Explicit Evaluation of the Metasurface Reactance and Susceptance, and the Scattered Fields}
\label{subsec:theory:metasurface_reactance_scattered_fields}
Equation \eqref{equ:TE_impedance_admittance_final} prescribes the required variation of the surface reactance and susceptance to implement the desirable HMS. However, in order to use this formula, we should instruct how the phases of the fields at $x\rightarrow 0^{\pm}$, defined as $\varphi_{\pm}\left(z\right)$, are to be evaluated. 

To that end, we first indicate how to find the reflected and transmitted fields from the given source field and the requirement that the power is locally conserved and the impedance is locally \textcolor{black}{equalized}. Local impedance \textcolor{black}{equalization} \eqref{equ:TE_local_impedance_matching_explicit} requires 
\begin{eqnarray}
&&\!\!\!\!\!\!E_y^{\mathrm{inc}}\left(0,z\right)-\frac{\eta}{\cos\theta_0}H_z^{\mathrm{inc}}\left(0,z\right) = \nonumber \\
	&& = -\left[E_y^{\mathrm{ref}}\left(0,z\right)-\frac{\eta}{\cos\theta_0}H_z^{\mathrm{ref}}\left(0,z\right)\right],
\label{equ:TE_local_impedance_matching_explicit_2}
\end{eqnarray}
from which, using the spectral representation of the fields \eqref{equ:TE_spectral_domain_Ey}-\eqref{equ:TE_spectral_domain_Hz}, the reflection coefficient can be evaluated as
\begin{eqnarray}
\leftscriptspace{e}{}{\Gamma}\left(k_t\right)=\frac{k\cos\theta_0-\beta}{k\cos\theta_0+\beta}
\label{equ:TE_reflection_coefficient}
\end{eqnarray}
\emph{regardless} of the source field. Equation \eqref{equ:TE_reflection_coefficient} is merely a private case of the Fresnel reflection formula for an incident plane-wave at an angle $\arcsin\left(k_t/k\right)$ and a transmitted plane-wave at an angle of $\theta_0$, travelling in media with the same permittivity and permeability \cite{Chew1990}. Actually, it is a generalization of the reflection coefficient derived in \cite{Selvanayagam2013} for passive lossless HMSs excited by a plane-wave. However, in the general case considered herein, the source field consists of an infinite number of plane-waves, propagating in different directions. Therefore, the reflection coefficient of \eqref{equ:TE_reflection_coefficient} actually ensures \emph{individual} impedance \textcolor{black}{equalization} of each of the plane-waves in the source field spectrum \eqref{equ:TE_spectral_domain_Ey}-\eqref{equ:TE_spectral_domain_Hz} to that of a plane-wave propagating in region 2 with an angle of $\theta_0$ with respect to the $x$ axis.

\begin{table*}[!t]
\renewcommand{\arraystretch}{1.3}
\caption{Summary of Design Procedure for Passive Lossless Huygens Metasurface for Directive Radiation}
\label{tab:design_rules}
\centering
\begin{tabular}{c |l| l| l| l}
\hline \hline
No. & Step & Input & Output & Relevant Equations\\
\hline \hline
1 & \begin{tabular}{l} Local impedance \\ \textcolor{black}{equalization} \end{tabular} & 
	\begin{tabular}{l} Incident fields: $\vec{H}^{\mathrm{inc}}\left(x,z\right)$, 
			$\vec{E}^{\mathrm{inc}}\left(x,z\right)$ \\
		Transmission angle: $\theta_0$
		\end{tabular}  & 
	\begin{tabular}{l} Reflected fields: $\vec{H}^{\mathrm{ref}}\left(x,z\right)$, 
			$\vec{E}^{\mathrm{ref}}\left(x,z\right)$ \\
		Total fields at $x\rightarrow 0^{-}$: $\left|F^{-}\left(z\right)\right|
			e^{j\varphi_{-}\left(z\right)}$
		\end{tabular} &
	\begin{tabular}{l}
		\eqref{equ:TE_spectral_domain_Ey}-\eqref{equ:TE_spectral_domain_Hz}, \\
		 \eqref{equ:TE_local_impedance_matching_explicit_2}-\eqref{equ:TE_spectral_domain_F_minus}
	\end{tabular} 
	\\
\hline
2 & \begin{tabular}{l} Local power \\ conservation \end{tabular} & 
	\begin{tabular}{l} Field magnitude at $x\rightarrow 0^{-}$: $\left|F^{-}\left(z\right)\right|$ 
			 \\
		Transmission angle: $\theta_0$
		\end{tabular}  & 
	\begin{tabular}{l} Field magnitude at $x\rightarrow 0^{+}$: $W\left(0,z\right)$ 
			 \\
		Transmitted fields: $\vec{H}^{\mathrm{trans}}\left(x,z\right)$, 
			$\vec{E}^{\mathrm{trans}}\left(x,z\right)$ 
		\end{tabular} &
	\begin{tabular}{l}
		\eqref{equ:TE_window_function_explicit}-\eqref{equ:TE_transmission_spectral_domain}, \\
		 \eqref{equ:TE_spectral_domain_Ey}-\eqref{equ:TE_spectral_domain_Hz}, 
		 \eqref{equ:TE_window_function_Ey}-\eqref{equ:TE_window_function_Hz}
	\end{tabular} 
	\\
\hline
3 & \begin{tabular}{l} Metasurface \\ reactance \end{tabular} & 
	\begin{tabular}{l} Field phase at $x\rightarrow 0^{-}$: $\varphi_{-}\left(z\right)$ 
			 \\
		Transmission angle: $\theta_0$
		\end{tabular}  & 
	\begin{tabular}{l} Surface impedance: $Z_{se}\left(z\right)$ 
			 \\
		Surface admittance: $Y_{sm}\left(z\right)$ 
		\end{tabular} &
	\begin{tabular}{l}
		\eqref{equ:TE_dimensionless_quantities_passive_HMS}-\eqref{equ:transmitted_wave_impedance}, \\
		 \eqref{equ:TE_phase_variation_plus}-\eqref{equ:TE_impedance_admittance_final_explicit}
	\end{tabular} 
	\\
\hline \hline
\end{tabular}
\end{table*}

Equation \eqref{equ:TE_reflection_coefficient} enables evaluation of the reflected fields (everywhere in region 1) via \eqref{equ:TE_spectral_domain_Ey}-\eqref{equ:TE_spectral_domain_Hz}. Combined with the given incident field, we may then utilize \eqref{equ:TE_dimensionless_quantities} to evaluate $\leftscript{e}{}{F}_{\!\!E}^{-}\left(z\right)=\leftscript{e}{}{F}_{\!\!H}^{-}\left(z\right)=\leftscript{e}{}{F}^{-}\left(z\right)$. 
Explicitly, this is given by
\begin{eqnarray}
\leftscript{e}{}{F}^{-}\left(z\right) = \frac{1}{2\pi}\int\limits_{-\infty}^{\infty}\frac{dk_t}{k\cos\theta_0+\beta}
	\leftscript{e}{}{f}\left(k_t\right)e^{jk_tz},
\label{equ:TE_spectral_domain_F_minus}
\end{eqnarray}
from which the total dimensionless field magnitude $\left|\leftscript{e}{}{F}\left(z\right)\right|$ and phase $\varphi_{-}\left(z\right)$ on the lower facet of the HMS are assessed, following \eqref{equ:TE_dimensionless_quantities_passive_HMS}. 

The local power conservation condition \eqref{equ:TE_local_power_conservation_impedance_matched} indicates that the dimensionless field magnitudes must be continuous at each $z$ along the metasurface. This facilitates the evaluation of the virtual aperture window function $\leftscriptspace{e}{}{W\left(0,z\right)}$ via \eqref{equ:TE_dimensionless_quantities}, namely
\begin{eqnarray}
\leftscriptspace{e}{}{W\left(0,z\right)} = \left|\frac{1}{2\pi}\!\!\int\limits_{-\infty}^{\infty}\!\!
	\frac{dk_t}{k\cos\theta_0+\beta}\leftscript{e}{}{f}\left(k_t\right)e^{jk_tz}\right| e^{-j\xi_0},
\label{equ:TE_window_function_explicit}
\end{eqnarray}
where $\xi_0$ is an arbitrary (constant) phase shift which may be introduced to the transmitted fields without affecting neither the radiation directivity nor the reflected fields (similarly to \cite{Selvanayagam2014}). 

From the definition \eqref{equ:TE_window_function_Ey} and the spectral representation \eqref{equ:TE_spectral_domain_Ey}, the spectral content of the transmitted field can be evaluated via the inverse Fourier transform,
\begin{eqnarray}
\leftscriptspace{e}{}{T}\left(k_t\right)=2\beta
\displaystyle\int\limits_{-\infty}^{\infty} \leftscriptspace{e}{}{W\left(0,z\right)}e^{-j\left(k_t+k\sin\theta_0\right)z}dz,
\label{equ:TE_transmission_spectral_domain}
\end{eqnarray}
using which the transmitted fields everywhere in region 2 may be computed from \eqref{equ:TE_spectral_domain_Ey}-\eqref{equ:TE_spectral_domain_Hz}.

Finally, the transmitted field phase variation on the HMS can be explicitly formulated using \eqref{equ:TE_window_function_explicit} and \eqref{equ:TE_dimensionless_quantities} as
\begin{eqnarray}
\varphi_{+}\left(z\right)=-kz\sin\theta_0-\xi_0
\label{equ:TE_phase_variation_plus}
\end{eqnarray}
\emph{regardless} of the source fields, and \eqref{equ:TE_impedance_admittance_final} can be rewritten as
\begin{eqnarray}
\left\lbrace
\begin{array}{l}
\vspace{5pt}
\!\!\!\!\leftscript{e}{}{Z}_{se}\left(z\right)=-j\dfrac{\leftscript{e}{}{Z}_0}{2}
	\cot\left[\dfrac{kz\sin\theta_0+\xi_0+\angle\leftscript{e}{}{F}^{-}\left(z\right)}{2}\right]
	 \\
\!\!\!\!\leftscriptspace{e}{}{Y}_{sm}\left(z\right)=-j\dfrac{\leftscriptspace{e}{}{Y}_0}{2}
	\cot\left[\dfrac{kz\sin\theta_0+\xi_0+\angle\leftscript{e}{}{F}^{-}\left(z\right)}{2}\right],
\end{array}\!\!\!\!\!
\right.
\label{equ:TE_impedance_admittance_final_explicit}
\end{eqnarray}
where $\angle\leftscript{e}{}{F}^{-}\left(z\right)=\varphi_{-}\left(z\right)$ is the phase of the dimensionless field parameter given by \eqref{equ:TE_spectral_domain_F_minus}.

The design procedure described in this Section is summarized in Table \ref{tab:design_rules}, with references to relevant equation numbers in the text. \textcolor{black}{It should be noted that the fields used to design the HMS are approximate, valid subject to the slowly-varying envelope condition \eqref{equ:TE_slowly_varying_W}; thus, after the design procedure is completed, and the predicted transmitted fields have been evaluated via \eqref{equ:TE_transmission_spectral_domain}, the satisfaction of \eqref{equ:TE_slowly_varying_W} should be verified to assess the consistency of the theoretical derivation (See, e.g., Subsections \ref{subsec:results:plane_wave},\ref{subsubsec:results:line_source:results}). 
}

\section{Results and Discussion}
\label{sec:results}
\begin{figure*}[!t]
\centering
\includegraphics[width=18cm]{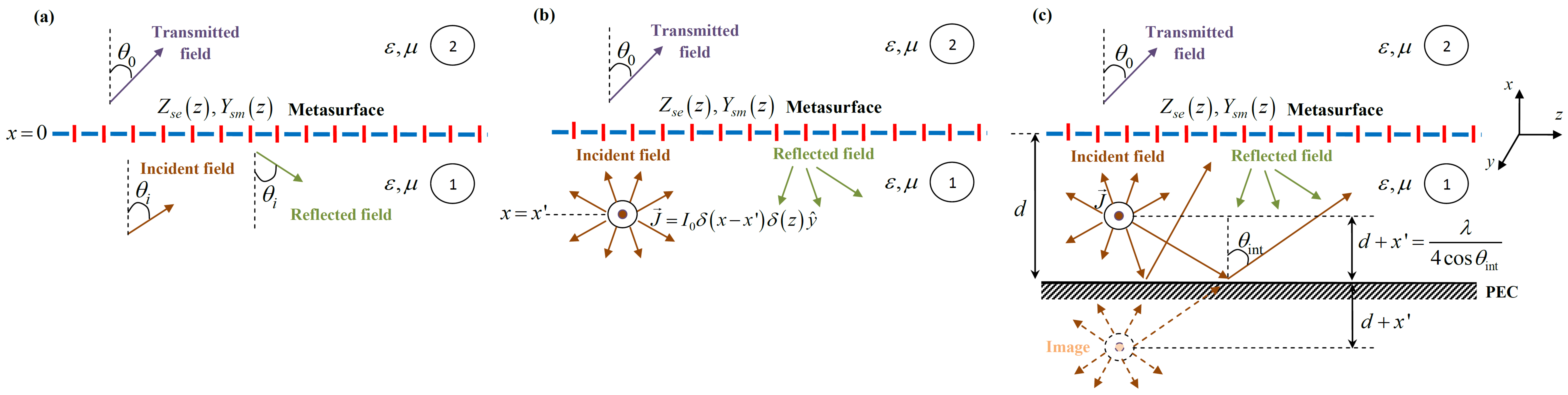}
\caption{Physical configurations of Huygens metasurfaces excited by (a) a plane wave forming an angle of $\theta_i$ with the $x$-axis (Subsection \ref{subsec:results:plane_wave}); (b) an electric line source situated at $\left(x,z\right)=\left(x',0\right)$ carrying a current of $\vec{J}=I_0\delta\left(x-x'\right)\delta\left(z\right)\hat{y}$ (Subsection \ref{subsec:results:line_source}); (c) the same electric line source positioned in front of a PEC, separated by $d$ from the HMS; $\theta_{\mathrm{int}}$ is the angle of internal constructive interference \textcolor{black}{between the source and its image} (Subsection \ref{subsec:results:line_source_PEC}). The transmitted field radiates towards the direction defined by $\theta_0$.}
\label{fig:PhysicalConfigurationThreeSources}
\end{figure*}

To verify our theory, we follow the design procedure outlined in Section \ref{sec:theory} and summarized in Table \ref{tab:design_rules} to design passive lossless HMS directive radiators for three different (TE-polarized) source excitations: a plane-wave, an electric line source (ELS), and an electric line source positioned in front of a perfect electric conductor (PEC) infinite plane (Fig. \ref{fig:PhysicalConfigurationThreeSources}). In the following Subsections we derive the expressions for the required surface impedance and surface admittance, and compare the performance of the HMS, predicted by semi-analytical means, to results of respective finite-element numerical simulations, in which the HMS is implemented using loaded loops and wires (similarly to \cite{Selvanayagam2013}). 

\subsection{Excitation by a Plane-wave}
\label{subsec:results:plane_wave}

We begin by verifying the consistency of our theory with previous HMS design formulae derived in \cite{Selvanayagam2013} for the private case of plane-wave excitation. This would also serve as a simple demonstration of the proposed design procedure, with indication of the relevant steps in Table \ref{tab:design_rules}.

In the configuration under consideration, the source field is produced by a TE-polarized plane-wave travelling at an angle $\theta_i$ with respect to the $x$-axis incident upon the HMS (Fig. \ref{fig:PhysicalConfigurationThreeSources}(a)). The spectrum of the source field is thus given by
\begin{eqnarray}
\leftscriptspace{e}{}{f}\left(k_t\right)=4\pi\beta\delta\left(k_t-k_{t,i}\right),
\label{equ:plane_wave_source_spectrum}
\end{eqnarray}
where $k_{t,i}=-k\sin\theta_i$. To enforce local impedance \textcolor{black}{equalization} (Table \ref{tab:design_rules} / Step 1), the reflection coefficient is defined according to \eqref{equ:TE_reflection_coefficient}, using which the incident and reflected fields \eqref{equ:TE_spectral_domain_Ey}-\eqref{equ:TE_spectral_domain_Hz} may be evaluated
\begin{eqnarray}
&&\left\lbrace
\begin{array}{l}
\!\!\!\!E_y^{\mathrm{inc}}\left(x,z\right) = k\eta I_0 e^{-jkx\cos\theta_i}e^{-jkz\sin\theta_i} \\
\!\!\!\!E_y^{\mathrm{ref}}\left(x,z\right) = -\dfrac{\cos\theta_0-\cos\theta_i}{\cos\theta_0+\cos\theta_i} E_y^{\mathrm{inc}}\left(x,z\right)
\end{array}\!\!\!\!
\right.
\label{equ:plane_wave_incident_reflected_Ey} \\
&&\left\lbrace
\begin{array}{l}
\!\!\!\!H_z^{\mathrm{inc}}\left(x,z\right) = k I_0\cos\theta_i e^{-jkx\cos\theta_i}e^{-jkz\sin\theta_i} \\
\!\!\!\!H_z^{\mathrm{ref}}\left(x,z\right) = \dfrac{\cos\theta_0-\cos\theta_i}{\cos\theta_0+\cos\theta_i}H_z^{\mathrm{inc}}\left(x,z\right),
\end{array}\!\!\!\!
\right.
\label{equ:plane_wave_incident_reflected_Hz}
\end{eqnarray}
and it is readily verified that the fields indeed satisfy \eqref{equ:TE_local_impedance_matching_explicit_2}. The dimensionless total field at $x\rightarrow 0^{-}$ is thus given by \eqref{equ:TE_spectral_domain_F_minus}
\begin{eqnarray}
\leftscript{e}{}{F}^{-}\left(z\right) = \frac{2\cos\theta_i}{\cos\theta_0+\cos\theta_i}
	e^{-jkz\sin\theta_i},
\label{equ:plane_wave_F_minus}
\end{eqnarray}
from which we arrive at the desirable output of [Table \ref{tab:design_rules} / Step 1], namely,
\begin{eqnarray}
\left|\leftscript{e}{}{F}^{-}\left(z\right)\right| \equiv \frac{2\cos\theta_i}{\cos\theta_0+\cos\theta_i}, & 
	\varphi_{-}\left(z\right)=-kz\sin\theta_i,
\label{equ:plane_wave_F_minus_mag_phase}
\end{eqnarray}
and we note that $\theta_i,\theta_0\in\left(-\pi/2,\pi/2\right)$.

Local power conservation (Table \ref{tab:design_rules} / Step 2) essentially requires that the magnitude of the virtual aperture window function on the upper facet of the HMS would follow the magnitude of the total fields (incident+reflected) on its lower facet. Hence, $W\left(0,z\right)$ is given by combining  \eqref{equ:TE_window_function_explicit} with \eqref{equ:plane_wave_F_minus_mag_phase}, yielding
\begin{eqnarray}
\leftscriptspace{e}{}{W\left(0,z\right)} \equiv \frac{2\cos\theta_i}{\cos\theta_0+\cos\theta_i}e^{-j\xi_0},
\label{equ:plane_wave_window_function}
\end{eqnarray}
where $\xi_0$ is an arbitrary phase, in case any is desirable. Using the last result and \eqref{equ:TE_transmission_spectral_domain}, the plane-wave spectrum of the transmitted wave can be evaluated,
\begin{eqnarray}
\leftscriptspace{e}{}{T}\left(k_t\right)=4\pi\beta\delta\left(k_t-k_{t,0}\right) \frac{2\cos\theta_i}{\cos\theta_0+\cos\theta_i}e^{-j\xi_0},
\label{equ:plane_wave_transmission_spectrum}
\end{eqnarray}
and subsequently, from \eqref{equ:TE_spectral_domain_Ey}-\eqref{equ:TE_spectral_domain_Hz}, also the transmitted fields, 
\begin{eqnarray}
&&\!\!\!\!\!\!\!\!E_y^{\mathrm{trans}}\left(x,z\right) = k\eta I_0 \frac{2\cos\theta_ie^{-j\xi_0}}{\cos\theta_0+\cos\theta_i}
	e^{-jkx\cos\theta_0}e^{-jkz\sin\theta_0}
\nonumber \\
&&\!\!\!\!\!\!\!\!H_z^{\mathrm{trans}}\left(x,z\right) = \frac{\cos\theta_0}{\eta}E_y^{\mathrm{trans}}\left(x,z\right).
\label{equ:plane_wave_transmitted_fields}
\end{eqnarray}
This completes Step 2 of the design procedure.

Finally, we may use \eqref{equ:TE_impedance_admittance_final_explicit} with the phase calculated in \eqref{equ:plane_wave_F_minus_mag_phase} to define the surface impedance and surface admittance required to implement the desirable HMS (Table \ref{tab:design_rules} / Step 3). These are given by
\begin{eqnarray}
\left\lbrace
\begin{array}{l}
\vspace{5pt}
\!\!\!\!\leftscript{e}{}{Z}_{se}\left(z\right)=-j\dfrac{\leftscript{e}{}{Z}_0}{2}
	\cot\left[\dfrac{kz\left(\sin\theta_0-\sin\theta_i\right)+\xi_0}{2}\right]
	 \\
\!\!\!\!\leftscriptspace{e}{}{Y}_{sm}\left(z\right)=-j\dfrac{\leftscriptspace{e}{}{Y}_0}{2}
	\cot\left[\dfrac{kz\left(\sin\theta_0-\sin\theta_i\right)+\xi_0}{2}\right],
\end{array}\!\!\!\!\!
\right.
\label{equ:plane_wave_impedance_admittance_final}
\end{eqnarray}
which, when $\xi_0=0$, coincide with the results obtained in \cite{Selvanayagam2013} for the same configuration (recall $\leftscript{e}{}{Z}_0\!=\!1/\leftscriptspace{e}{}{Y}_0\!=\!\eta/\cos\theta_0$). 

As this type of HMS was thoroughly investigated in \cite{Selvanayagam2013,Selvanayagam2013_1}, including demonstration of its performance using numerical simulation tools, we would not discuss it here further. It is, however, worthwhile to note that in the context of the design procedure formulated in Section \ref{sec:theory}, the virtual aperture window function $\leftscriptspace{e}{}{W\left(x,z\right)}$ formed by the HMS has infinite extent in this case. More precisely, from \eqref{equ:plane_wave_transmitted_fields} and \eqref{equ:TE_window_function_Ey} it follows that $\leftscriptspace{e}{}{W\left(x,z\right)}$ is constant. Therefore, the satisfaction of \eqref{equ:TE_slowly_varying_W} is trivial for any transmission angle (${\left|\theta_0\right|\!<\!\pi/2}$), indicating that the theoretical prediction of the HMS functionality is valid.   

\subsection{Excitation by an Electric Line Source}
\label{subsec:results:line_source}

Next, we consider the scenario in which the HMS is excited by an electric line source $\vec{J}=I_0\delta\left(x-x'\right)\delta\left(z\right)\hat{y}$ situated in region 1, where according to our convention $x'<0$ (Fig. \ref{fig:PhysicalConfigurationThreeSources}(b)). This configuration is different from the plane-wave excitation scenario discussed in Subsection \ref{subsec:results:plane_wave} in three significant aspects, all originate from the localized nature of the source: first, the source introduces discontinuity to the fields at $x=x'$ (region 1); second, its spectral representation consists of a wide range of plane-waves; third, its illumination of the HMS is non-uniform, creating a localized spot on the virtual aperture. In the following Subsubsections, we will emphasize the effects of these differences on the design procedure.

\subsubsection{HMS Design}
\label{subsubsec:results:line_source:design}
As mentioned in the previous paragraph, the derivation of the source plane-wave spectrum $\leftscript{e}{}{f}\left(k_t\right)$ now involves a source condition\footnote{This would be a consequence of the introduction of a (singular) nonhomogeneous term to the Helmholtz (wave) equation of \eqref{equ:TE_Maxwell_equations}, required if the whole region 1 ($x<0$) is to be described by this equation.
}, requiring the discontinuity of the derivative of the characteristic Green's function at $x=x'$ \cite{FelsenMarcuvitz1973,Chew1990,Holloway2012_1,Epstein2013_3}. Moreover, to satisfy the radiation condition at $x\rightarrow -\infty$ we should use a different $x$ dependency for the incident field in the region $x<x'$ than we used for the source-free region $x'<x<0$, manifesting the fact that the plane-waves propagate away from the source at all regions. 

Considering these conditions, the incident and reflected fields in the whole region 1 ($x<0$) may be formulated as
\begin{eqnarray}
&&\!\!\!\!\!\!\!\!\!\!\!\!\left\lbrace
\begin{array}{l}
\!\!\!\!E_y^{\mathrm{inc}}\left(x,z\right) = k\eta \dfrac{I_0}{2\pi}\displaystyle\int\limits_{-\infty}^{\infty}\dfrac{dk_t}{2\beta}
	e^{-j\beta \left|x-x'\right|}e^{jk_tz} \\
\!\!\!\!E_y^{\mathrm{ref}}\left(x,z\right) = -k\eta \dfrac{I_0}{2\pi}\displaystyle\int\limits_{-\infty}^{\infty}\dfrac{dk_t}{2\beta}
	\leftscriptspace{e}{}{\Gamma}\left(k_t\right)e^{j\beta \left(x+x'\right)}e^{jk_tz}
\end{array}\!\!\!\!
\right.
\label{equ:line_source_incident_reflected_Ey} \\
&&\!\!\!\!\!\!\!\!\!\!\!\!\left\lbrace
\begin{array}{l}
\!\!\!\!H_z^{\mathrm{inc}}\left(x,z\right) = \pm \dfrac{I_0}{2\pi}\displaystyle\int\limits_{-\infty}^{\infty}\dfrac{dk_t}{2}
	 e^{-j\beta \left|x-x'\right|}e^{jk_tz} \\
\!\!\!\!H_z^{\mathrm{ref}}\left(x,z\right) = \dfrac{I_0}{2\pi}\displaystyle\int\limits_{-\infty}^{\infty}\dfrac{dk_t}{2}
	\leftscriptspace{e}{}{\Gamma}\left(k_t\right)e^{j\beta \left(x+x'\right)}e^{jk_tz},
\end{array}\!\!\!\!
\right.
\label{equ:line_source_incident_reflected_Hz}
\end{eqnarray}
where the reflection coefficient remains a free parameter and the upper and lower signs in \eqref{equ:line_source_incident_reflected_Hz} should be used when ${x>x'}$ or ${x<x'}$, respectively; this change of signs establishes the required discontinuity in the magnetic field at the source position. Comparing \eqref{equ:line_source_incident_reflected_Ey}-\eqref{equ:line_source_incident_reflected_Hz} in the \emph{source-free region} $x'<x<0$ with the general form \eqref{equ:TE_spectral_domain_Ey}-\eqref{equ:TE_spectral_domain_Hz} yields the expression for the source-related plane-wave spectrum, namely
\begin{eqnarray}
\leftscriptspace{e}{}{f}\left(k_t\right)=e^{j\beta x'},
\label{equ:line_source_source_spectrum}
\end{eqnarray}
which forms the necessary input to begin our design procedure. 

To enforce local impedance \textcolor{black}{equalization} (Table \ref{tab:design_rules} / Step 1), we define the reflection coefficient according to \eqref{equ:TE_reflection_coefficient}, which then enables the evaluation of the incident and reflected fields across the entire region 1. The dimensionless total field at $x\rightarrow 0^{-}$ is thus given by \eqref{equ:TE_spectral_domain_F_minus}
\begin{eqnarray}
\leftscript{e}{}{F}^{-}\left(z\right) = \frac{1}{2\pi}\int\limits_{-\infty}^{\infty}\frac{dk_t}{k\cos\theta_0+\beta}
	e^{j\beta x'}e^{jk_tz}.
\label{equ:line_source_spectral_domain_F_minus}
\end{eqnarray}

As a result of the wide spectral content of the source, 
and the fact that the reflection coefficient required to guarantee local impedance \textcolor{black}{equalization} varies with the transverse wavenumber $k_t$, obtaining an analytical closed-form expression for $\leftscript{e}{}{F}^{-}\left(z\right)$ is not trivial as it was for the plane-wave excitation scenario (See \eqref{equ:plane_wave_F_minus}). Nonetheless, as the integral of \eqref{equ:line_source_spectral_domain_F_minus} consists of a slowly varying part $1/\left(k\cos\theta_0+\beta\right)$ and an oscillatory part $e^{j\beta x'}e^{jk_tz}$, we may employ asymptotic evaluation techniques (e.g., the steepest-descent-path method \cite{FelsenMarcuvitz1973}) to evaluate it in closed-form for those evaluation points (on the metasurface) which are in the far field of the source. If, however, the distance between the evaluation point (on the metasurface) and the source $\rho'=\sqrt{x'^2+z^2}$ is not very large with respect to the wavelength, the oscillatory part variation is moderate enough such that, in general, $\leftscript{e}{}{F}^{-}\left(z\right)$ may be evaluated by straightforward numerical integration\footnote{In fact, the integrand is bounded on the entire real $k_t$ axis due to the reflection coefficient.}. 

One way or the other, the magnitude and phase functions ($\left|\leftscript{e}{}{F}^{-}\left(z\right)\right|$ and $\varphi_{-}\left(z\right)$, respectively) can be calculated from \eqref{equ:line_source_spectral_domain_F_minus}, as required in [Table \ref{tab:design_rules} / Step 1]. These would determine, respectively, the profile of the virtual aperture window function (following \eqref{equ:TE_window_function_explicit} [Table \ref{tab:design_rules} / Step 2]), and the phase compensation required by the HMS to ensure the fields on that aperture carry a linear phase (following \eqref{equ:TE_phase_variation_plus} [Table \ref{tab:design_rules} / Step 3]). 

Finally, completing Step 2 and Step 3 of the design procedure, we are able to evaluate the fields at each point in space, as well as the desirable surface impedance and surface admittance defining the HMS. The latter may be written as a generalized form of \eqref{equ:plane_wave_impedance_admittance_final}, namely,
\begin{eqnarray}
\left\lbrace
\begin{array}{l}
\vspace{5pt}
\!\!\!\!\leftscript{e}{}{Z}_{se}\left(z\right)=-j\dfrac{\leftscript{e}{}{Z}_0}{2}
	\cot\left[\dfrac{kz\left(\sin\theta_0-\sin\theta_i\left(z\right)\right)+\xi_0}{2}\right]
	 \\
\!\!\!\!\leftscriptspace{e}{}{Y}_{sm}\left(z\right)=-j\dfrac{\leftscriptspace{e}{}{Y}_0}{2}
	\cot\left[\dfrac{kz\left(\sin\theta_0-\sin\theta_i\left(z\right)\right)+\xi_0}{2}\right],
\end{array}\!\!\!\!\!
\right.
\label{equ:line_source_impedance_admittance_final}
\end{eqnarray}
where we used the definition of the equivalent angle of incidence $\theta_i\left(z\right)$
\begin{eqnarray}
	\varphi_{-}\left(z\right)=-kz\sin\theta_i\left(z\right)
\label{equ:equivalent_angle_incidence}
\end{eqnarray}
in analogy to \eqref{equ:plane_wave_F_minus_mag_phase}.
As in Subsection \ref{subsec:results:plane_wave}, upon evaluation of the transmitted fields, the satisfaction of the slowly-varying \textcolor{black}{envelope} condition \eqref{equ:TE_slowly_varying_W} should be verified to ensure the consistency of the design procedure.

\subsubsection{Virtual Aperture Engineering}
\label{subsubsec:results:line_source:aperture}
Before we proceed to demonstrate the performance of several ELS-excited HMS designs, we refer to the last point mentioned in the opening paragraph of this Subsection. As part of the device engineering, it is important to control the shape of the virtual aperture, as it determines to a large extent the width and directivity of the transmitted radiation. However, as opposed to the case of plane-wave excitation, in which the metasurface was illuminated uniformly across the entire $z$ axis, when finite-energy sources are used, the nature of the resultant virtual aperture is not as easily predicted. The reason for that is that the profile of the virtual aperture window function is not constant anymore, and is determined by the total field at $x\rightarrow 0^{-}$, i.e. the \emph{sum} of the incident and reflected fields; while the former is known, the latter is an outcome of the integration of the individually reflected source plane-waves \eqref{equ:TE_reflection_coefficient}. 

Nevertheless, if the source is not illuminating the metasurface at a grazing angle, the reflection coefficient varies rather moderately with $k_t$ in the spectral region contributing dominantly to the reflected field. In that case, a good zero-order approximation for the shape of the virtual aperture window would be given by the magnitude of the incident field, thus providing a starting point for selecting sources suitable for a desirable virtual aperture design. More than that, the phase of the incident field will be then a reasonable approximation to $\angle\leftscript{e}{}{F}^{-}\left(z\right)$, providing an insight on the variation of $Z_{se}$ and $Y_{sm}$ along the HMS \eqref{equ:TE_impedance_admittance_final_explicit}. If, in addition, the evaluation point on the metasurface is in the far-field of the source, the equivalent angle of incidence $\theta_i\left(z\right)$ defined in \eqref{equ:equivalent_angle_incidence} receives an elegant physical interpretation: this is the angle of incidence of the ray ("local" plane-wave) incident upon the metasurface in the neighbourhood of $z$. Accordingly, \eqref{equ:line_source_impedance_admittance_final} can be interpreted as a generalization of \eqref{equ:plane_wave_impedance_admittance_final}, where due to the localized source, different points along the metasurface interact with plane-waves having different angles of incidence. 

\subsubsection{Numerical and Semi-analytical Results}
\label{subsubsec:results:line_source:results} 
\begin{figure*}[!t]
\centering
\includegraphics[width=15.5cm]{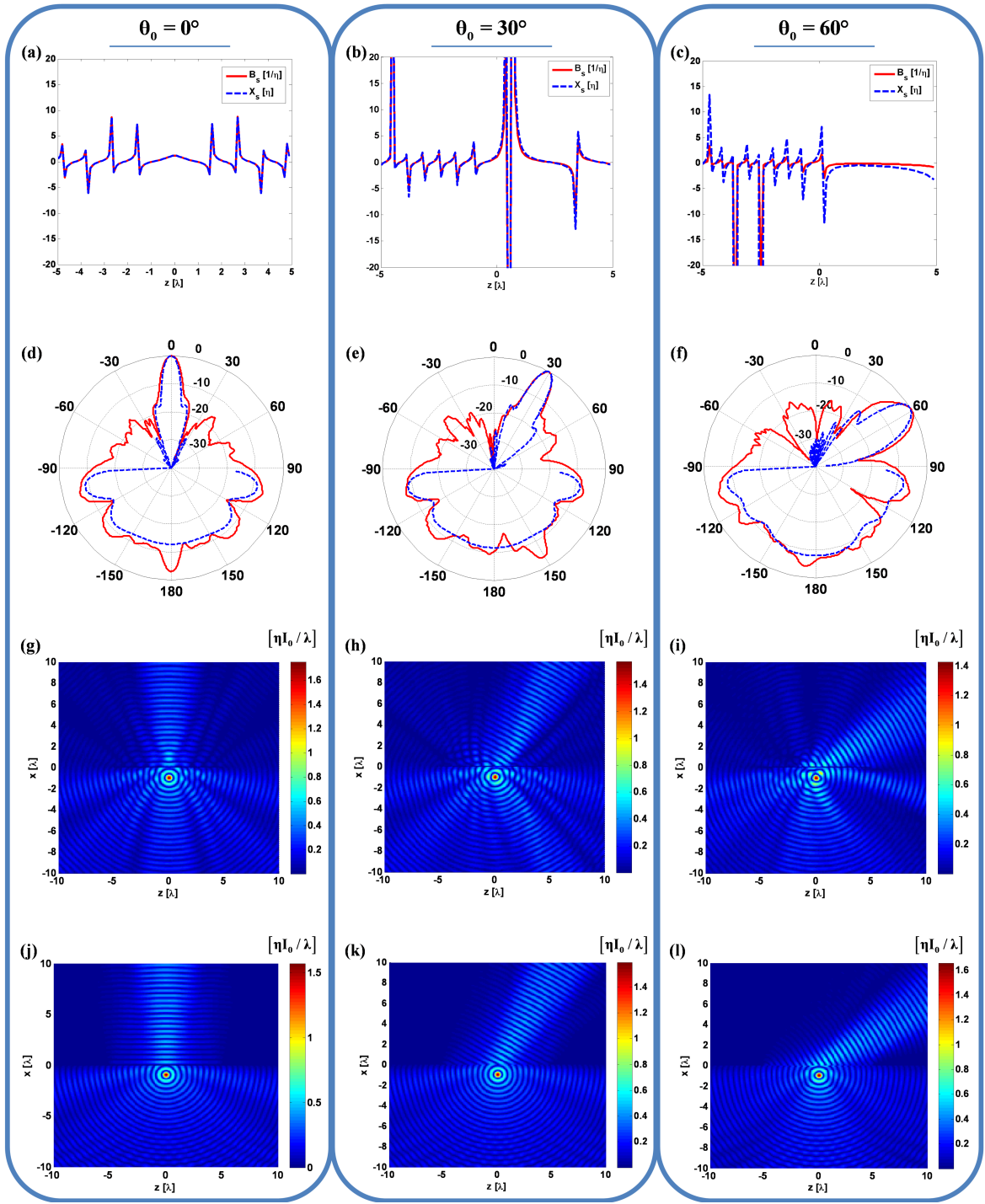}
\caption{Comparison between the theoretically (semi-analytical) predicted performance and the results of HFSS simulations of Huygens metasurfaces designed to convert the fields produced by an electric line source situated at $x'=-\lambda$ to directive radiation towards (a,d,g,j) $\theta_0=0^{\circ}$, (b,e,h,k) $\theta_0=30^{\circ}$, and (c,f,i,l) $\theta_0=60^{\circ}$. \textcolor{black}{(a)-(c)} Required surface reactance $X_s\left(z\right)=\Im\left\lbrace Z_{se}\left(z\right)\right\rbrace$ (blue dashed line) and surface susceptance $B_s\left(z\right)=\Im\left\lbrace Y_{sm}\left(z\right)\right\rbrace$ (red solid line) calculated from \eqref{equ:line_source_impedance_admittance_final}. \textcolor{black}{(d)-(f)} Theoretically predicted (blue dashed line) and HFSS-simulated (red solid lines) normalized radiation pattern (dB scale). \textcolor{black}{(g)-(i)} Real part of the electric field phasor $\left|\Re\left\lbrace E_y\left(x,z\right)\right\rbrace\right|$ as simulated by HFSS. (j)-(l) Theoretical prediction of $\left|\Re\left\lbrace E_y\left(x,z\right)\right\rbrace\right|$ (Appendix \ref{app:asymptotic_evaluation}).
}
\label{fig:LineSourceResults}
\end{figure*}

\begin{table*}[!t]
\begin{threeparttable}[b]
\renewcommand{\arraystretch}{1.3}
\caption{Performance Parameters of Line Source Excited HMS Directive Radiators Corresponding to Fig. \ref{fig:LineSourceResults}}
\label{tab:line_source_HMS_performance}
\centering
\begin{tabular}{l|c c c|c c c|c c c}
\hline \hline
& \multicolumn{3}{c}{$\theta_0=0^{\circ}$} & \multicolumn{3}{|c}{$\theta_0=30^{\circ}$} 
	& \multicolumn{3}{|c}{$\theta_0=60^{\circ}$} \\
\hline
& HFSS & Theory & \begin{tabular}{c} Relative \\ Performance \end{tabular}
& HFSS & Theory & \begin{tabular}{c} Relative \\ Performance \end{tabular}
& HFSS & Theory & \begin{tabular}{c} Relative \\ Performance \end{tabular} \\ 
\hline \hline \\[-1.3em]
	\begin{tabular}{l} Transmission Efficiency \end{tabular}
	 & $33\%$ & $42\%$ & $79\%$
	 & $36\%$ & $43\%$ & $84\%$
	 & $39\%$ & $42\%$ & $93\%$
	 \\	 \hline 
	 \begin{tabular}{l} Half-Power Beam Width \end{tabular}
	 & $8.4^{\circ}$ & $7.1^{\circ}$ & $85\%$
	 & $8.3^{\circ}$ & $7.9^{\circ}$ & $95\%$
	 & $16.8^{\circ}$ & $12^{\circ}$ & $71\%$
	 	 	\\	\hline 
	 \begin{tabular}{l} Aperture Efficiency \end{tabular}
	 & $60\%$ & $71\%$ & $85\%$
	 & $71\%$ & $74\%$ & $95\%$
	 & $57\%$ & $80\%$ & $71\%$
	 	 	 \\ \hline 
	 \begin{tabular}{l} Peak Directivity \end{tabular}
	 & $10.4$ & $19.2$ & $54\%$
	 & $11.4$ & $17.2$ & $66\%$
	 & $7.1$ & $11.1$ & $64\%$
	 	 	 	 \\		 	
\hline \hline
\end{tabular}
\end{threeparttable}
\end{table*}

To verify our formulation, we have designed and simulated three Huygens metasurfaces according to the procedure described in this Subsection (with $\xi_0=0$), designated to convert the fields produced by an electric line source at $x=x'=-\lambda$ to directive radiation towards $\theta_0=0^{\circ}$, $\theta_0=30^{\circ}$, and $\theta_0=60^{\circ}$ (Fig. \ref{fig:LineSourceResults}). 

For an HMS stretching from $z=-5\lambda$ to $z=5\lambda$ (total length $L=10\lambda$) the required variation of the surface reactance $X_s\left(z\right)=\Im\left\lbrace Z_{se}\left(z\right)\right\rbrace$ and surface susceptance $B_s\left(z\right)=\Im\left\lbrace Y_{sm}\left(z\right)\right\rbrace$ with $z$ is presented in Fig. \ref{fig:LineSourceResults}(a)-(c). For clarity, only values of $\left|X_s\right|<20\eta$ and $\left|B_s\right|<20/\eta$ are presented. Recalling the physical meaning of the equivalent angle of incidence \eqref{equ:equivalent_angle_incidence} we can indeed observe a change of signs in the cotangent argument of \eqref{equ:line_source_impedance_admittance_final} when the angles formed between the source and the evaluation point approaches $\theta_0$, i.e. in the proximity of $z/\left|x'\right|=\tan\theta_0$. This would happen around $z=0$ (Fig. \ref{fig:LineSourceResults}(a)), $z=0.577\lambda$ (Fig. \ref{fig:LineSourceResults}(b)), and $z=1.732\lambda$ (Fig. \ref{fig:LineSourceResults}(c)), for $\theta_0=0^{\circ}$, $\theta_0=30^{\circ}$, and $\theta_0=60^{\circ}$, respectively. 

Moreover, as the difference between the equivalent angles and the transmission angle becomes larger, the effective period of the cotangent should become shorter \eqref{equ:line_source_impedance_admittance_final}; indeed, if we focus on the $z<0$ region of the metasurface, where the equivalent angles of incidence are mostly negative, we observe that the effective period for $\theta_0=0^{\circ}$ (Fig. \ref{fig:LineSourceResults}(a)), is longer than that of $\theta_0=30^{\circ}$ (Fig. \ref{fig:LineSourceResults}(b)), which is, in turn, longer than the one corresponding to $\theta_0=60^{\circ}$ (Fig. \ref{fig:LineSourceResults}(c)). The practical implication of this observation is that if we desire to harness those rays incident upon the metasurface at angles which differ significantly from the desirable transmission angle, we should anticipate a fast variation of the corresponding surface impedance and admittance, which, in turn, requires smaller distances between the elements implementing the metasurface.

We have implemented the designed HMSs in a commercially available finite-element solver (ANSYS HFSS) using one hundred $\lambda/10$-long unit cells comprised of loaded wires and loops \cite{Selvanayagam2013, Selvanayagam2013_1}, as described in Appendix \ref{app:HFSS_implementation}. The HMS was excited by an electric line source carrying $I_0=1\mathrm{A}$ current oscillating at a frequency of $f=1.5\mathrm{GHz}$; the simulated electric field variation as a function of position, $\left|\Re\left\lbrace E_y\left(x,z\right)\right\rbrace\right|$, is presented in Fig. \ref{fig:LineSourceResults}(g)-(i) for the three HMSs considered. 

To compare these results with the theoretical predictions, we have calculated the spectral integrals \eqref{equ:TE_spectral_domain_Ey}-\eqref{equ:TE_spectral_domain_Hz} in conjunction with \eqref{equ:TE_reflection_coefficient},\eqref{equ:TE_window_function_explicit},\eqref{equ:TE_transmission_spectral_domain}, and \eqref{equ:line_source_incident_reflected_Ey}-\eqref{equ:line_source_source_spectrum}, to evaluate the fields in the region $\left(x,z\right)\in\left(-10\lambda,10\lambda\right)\times\left(-10\lambda,10\lambda\right)$, as presented in Fig. \ref{fig:LineSourceResults}(j)-(l). The theoretical plots rely on semi-analytical approximations (Appendix \ref{app:asymptotic_evaluation}), which assume the HMS is of infinite extent to calculate the fields in region 1 ($x<0$) and on the virtual aperture $x\rightarrow 0^+$. Then, to account for the finite HMS length when evaluating the fields in region 2, the virtual aperture window function $W\left(0,z\right)$ is truncated at $z=\pm L/2$ before utilizing \eqref{equ:TE_transmission_spectral_domain}. These approximations yield accurate results when most of the excitation power interacts with the HMS on its finite extent ($\left|z\right|<L/2$). 

In addition, the steepest-descent-path method is employed to evaluate the radiated power in the far-field regions ${x\rightarrow\pm\infty}$ using the spectral integrals \eqref{equ:TE_spectral_domain_Ey}-\eqref{equ:TE_spectral_domain_Hz} and based on the same assumptions \cite{FelsenMarcuvitz1973,Chew1990,Epstein2013_3} (Appendix \ref{app:asymptotic_evaluation}). 
This allows us to compare in Fig. \ref{fig:LineSourceResults}(d)-(f) the theoretically predicted far-field radiation patterns (dashed blue line) with the ones calculated by the HFSS simulation (solid red line); all radiation patterns are normalized to their maximum.

Although some discrepancies between the simulation results and the theoretical predictions are observed in Fig. \ref{fig:LineSourceResults}(d)-(l), it is clear that the designed HMS successfully convert the line source fields to directive radiation toward the desirable angle. Both the beam-width and the immediate side lobe levels are in a good agreement with the semi-analytical theory. In all three cases, the simulated directivity values outside the main beam (in region 2) are at least $15\mathrm{dB}$ below the peak directivity.

Importantly, the theoretical calculations indicate that the slowly-varying \textcolor{black}{envelope} condition \eqref{equ:TE_slowly_varying_W} is indeed satisfied for all transmission angles considered, except at some points towards the edge of the metasurface, where the exclusion of diffraction effects introduce a discontinuity in the fields along the $z$ axis (not shown). Another support for the validity of the theoretical results is provided by the fact that the absolute values of the fields as presented in Fig. \ref{fig:LineSourceResults}(g)-(i) and Fig. \ref{fig:LineSourceResults}(j)-(l) in $\eta I_0/\lambda$ units, are to scale.

Table \ref{tab:line_source_HMS_performance} concentrates performance parameters calculated from the theoretical and simulated radiation patterns of Fig. \ref{fig:LineSourceResults}(d)-(f). These include the transmission efficiency, i.e. the ratio between the power transmitted to region 2 and the total power radiated by the source; the half-power beamwidth (HPBW), i.e. the angular difference between the half-power points; the aperture efficiency, i.e. the ratio between the HPBW of the HMS radiation and the HPBW of a uniformly excited aperture with the same length $L$ \cite{Balanis1997}; and the (2D) peak directivity \cite{Lovat2006}. For each parameter, Table \ref{tab:line_source_HMS_performance} indicates its value as calculated from the simulation results and the theoretical predictions, and the respective relative performance, defined as the ratio between the two. In consistency with the results presented in Fig. \ref{fig:LineSourceResults}(d)-(l), the performance parameters also indicate that the power radiated into region 2 is successfully funnelled into a directive beam, with performance comparable with the theoretical predictions, peak directivity excepted. 

In view of the theoretically predicted values themselves, two comments are in place. First, we refer to the predicted transmission efficiencies, which are calculated to be around $42\%$. Due to the finite length $L$ of the HMS, only \textcolor{black}{${\arctan\left(L/\left|2x'\right|\right)/\pi}$} of the line source power interacts with the HMS. If we assume minor reflections from the metasurface, and consider $L/\left|x'\right|=10$ as in our case, this rough estimation leads to a transmission efficiency of $44\%$, very close to the values in Table \ref{tab:line_source_HMS_performance}. Hence, these values indicate that if the HMS implementation would perfectly match its design, most of the power interacting with the metasurface is expected to be transmitted to region 2. In other words, the reflection coefficient resulting from the enforcement of local impedance \textcolor{black}{equalization}, of which we have little control, should not deteriorate significantly the HMS performance. 
  
The second comment refers to the aperture efficiencies presented in Table \ref{tab:line_source_HMS_performance}, which do not exceed $80\%$ for all transmission angles considered, even in the ideal theoretical scenario. The reason for these values is the utilization of local power conservation in the HMS design, that coerces the virtual aperture window function to follow the profile of the total (incident+reflected) fields at $x\rightarrow 0^{-}$. As the line source is situated only $\left|x'\right|=\lambda$ away from the metasurface in our configuration, most of the incident power is concentrated in a small region near $z=0$. Thus, effectively, the full length of the HMS cannot be utilized for radiation, and, in turn, the theoretical limit for the aperture efficiency is below $100\%$.   

To conclude this Subsection, we note that both the results presented in Table \ref{tab:line_source_HMS_performance} and in Fig. \ref{fig:LineSourceResults}(d)-(l) indicate that the main deviations from the theoretical predictions are the increased reflections from the metasurface in region 1 and the incomplete elimination of the incident field in region 2; both discrepancies contribute to a significant difference in peak directivity (Table \ref{tab:line_source_HMS_performance}). We believe that these differences originate in the fact that the unit cells implementing the HMS have not yet been optimized to exhibit the prescribed surface admittance and impedance accurately over the entire required dynamical range. However, the optimization of the metasurface unit cells requires specialized treatment, including scattering element selection \cite{Kuester2003, Tretyakov2003}, and is outside the scope of this paper. 

\subsection{Excitation by an Electric Line Source in front of a PEC}
\label{subsec:results:line_source_PEC}

To further illustrate the versatility of our formulation, we consider a third excitation configuration, that of an electric line source positioned in front of a PEC, the latter is separated from the HMS by a distance $d$ (Fig. \ref{fig:PhysicalConfigurationThreeSources}(c)); as in Subsection \ref{subsec:results:line_source}, the line source current is given by $\vec{J}=I_0\delta\left(x-x'\right)\delta\left(z\right)\hat{y}$. We present this configuration herein for two reasons: first, to demonstrate how scenarios including multiple reflections (more generally, plane-stratified configurations) can be treated using the design procedure presented in Section \ref{sec:theory}; second, to provide an example for virtual aperture engineering via careful selection of the source excitation.

\subsubsection{HMS Design}
\label{subsubsec:results:line_source_PEC:design}
When region 1 contains not only sources but also scatterers (e.g., abrupt interfaces) the field expressions must take into account the boundary conditions induced by these scatterers, on top of the source condition already encountered in Subsection \ref{subsec:results:line_source}. The enforcement of these boundary conditions gives rise to multiple-reflection terms in the spectral response of the fields \cite{FelsenMarcuvitz1973,Chew1990,Epstein2013_3}, which in turn form a dependency between the source-related spectrum $\leftscript{e}{}{f}\left(k_t\right)$ of \eqref{equ:TE_spectral_domain_Ey}-\eqref{equ:TE_spectral_domain_Hz} and the reflection coefficient of the HMS.

In the configuration considered herein (Fig. \ref{fig:PhysicalConfigurationThreeSources}(c)) the boundary conditions introduced by the scatterers at $x\le x'$ require that the tangential electric field vanishes on the PEC, i.e. $E_y\left(-d,z\right)=0$ for each $z$. Combining this requirement with the source condition at $x=x'$ yields the following expressions for the incident and reflected fields \eqref{equ:TE_spectral_domain_Ey}-\eqref{equ:TE_spectral_domain_Hz} \cite{FelsenMarcuvitz1973,Chew1990,Epstein2013_3}
\begin{eqnarray}
&&\!\!\!\!\!\!\!\!\!\!\!\!\!\!\!\!\left\lbrace
\begin{array}{l}
\vspace{5pt}
\!\!\!\!E_y^{\mathrm{inc}}\left(x,z\right) \!=\! k\eta \dfrac{I_0}{2\pi}
\!\!\!\displaystyle\int\limits_{-\infty}^{\infty}\!\!\!\dfrac{dk_t}{2\beta}\!\!
	\left\lbrace\!\!\!\!
	\begin{array}{l}
	\vspace{5pt}
	\dfrac{e^{-j\beta\left(d+x_<\right)}-e^{j\beta\left(d+x_<\right)}}
		{e^{j\beta d}-\leftscriptspace{e}{}{\Gamma}\left(k_t\right)e^{-j\beta d}}
	\\ e^{-j\beta x_>}e^{jk_tz}
	\end{array}\!\!\!\!
	\right\rbrace
	 \\
\!\!\!\!E_y^{\mathrm{ref}}\left(x,z\right) \!=\!  -k\eta \dfrac{I_0}{2\pi}
\!\!\!\displaystyle\int\limits_{-\infty}^{\infty}\!\!\!\dfrac{dk_t}{2\beta}\!\!
	\left\lbrace\!\!\!\!
	\begin{array}{l}
	\vspace{5pt}
	\dfrac{e^{-j\beta\left(d+x_<\right)}-e^{j\beta\left(d+x_<\right)}}
		{e^{j\beta d}-\leftscriptspace{e}{}{\Gamma}\left(k_t\right)e^{-j\beta d}}
	\\ \leftscriptspace{e}{}{\Gamma}\left(k_t\right)e^{j\beta x_>}e^{jk_tz}
	\end{array}\!\!\!\!
	\right\rbrace
\end{array}\!\!\!\!
\right.
\label{equ:line_source_PEC_incident_reflected_Ey} \\
&&\!\!\!\!\!\!\!\!\!\!\!\!\!\!\!\!\left\lbrace
\begin{array}{l}
\vspace{5pt}
\!\!\!\!H_z^{\mathrm{inc}}\left(x,z\right) \!=\! \dfrac{I_0}{2\pi}
\!\!\!\displaystyle\int\limits_{-\infty}^{\infty}\!\!\!\dfrac{dk_t}{2}\!\!
	\left\lbrace\!\!\!\!
	\begin{array}{l}
	\vspace{5pt}
	\dfrac{e^{-j\beta\left(d+x_<\right)}\mp e^{j\beta\left(d+x_<\right)}}
		{e^{j\beta d}-\leftscriptspace{e}{}{\Gamma}\left(k_t\right)e^{-j\beta d}}
	\\ e^{-j\beta x_>}e^{jk_tz}
	\end{array}\!\!\!\!
	\right\rbrace
	 \\
\!\!\!\!H_z^{\mathrm{ref}}\left(x,z\right) \!=\!  \dfrac{I_0}{2\pi}
\!\!\!\displaystyle\int\limits_{-\infty}^{\infty}\!\!\!\dfrac{dk_t}{2}\!\!
	\left\lbrace\!\!\!\!
	\begin{array}{l}
	\vspace{5pt}
	\dfrac{\pm e^{-j\beta\left(d+x_<\right)}-e^{j\beta\left(d+x_<\right)}}
		{e^{j\beta d}-\leftscriptspace{e}{}{\Gamma}\left(k_t\right)e^{-j\beta d}}
	\\ \leftscriptspace{e}{}{\Gamma}\left(k_t\right)e^{j\beta x_>}e^{jk_tz}
	\end{array}\!\!\!\!
	\right\rbrace
	
\end{array}\!\!\!\!
\right.
\label{equ:line_source_PEC_incident_reflected_Hz}
\end{eqnarray}
where $x_<=\min\left\lbrace x,x'\right\rbrace$, $x_>=\max\left\lbrace x,x'\right\rbrace$, and the upper and lower signs in \eqref{equ:line_source_PEC_incident_reflected_Hz} should be used when $x>x'$ or $x<x'$, respectively. As in \eqref{equ:line_source_incident_reflected_Hz}, this change of signs for the regions below and above the source provides the required discontinuity in the magnetic field at $x=x'$ due to the electric current (source condition). We emphasize once more that the formulation of \eqref{equ:line_source_PEC_incident_reflected_Ey}-\eqref{equ:line_source_PEC_incident_reflected_Hz} is valid for any reflection coefficient dependency $\leftscriptspace{e}{}{\Gamma}\left(k_t\right)$, retaining this degree of freedom required to employ our design procedure. 

The fraction in the braces of the integrands \eqref{equ:line_source_PEC_incident_reflected_Ey} accounts for the reflection from the PEC, and its numerator vanishes when $x=x_<=-d$ as required. Its denominator corresponds to the multiple reflections taking place between the HMS and the PEC\cite{Chew1990, Epstein2013_3}; consequently, the poles of the integrands correspond to guided or leaky modes of this structure \cite{FelsenMarcuvitz1973,Holloway2012_1}. Moreover, although the source-related spectrum is now dependent on the reflection coefficient, it can be readily verified that when $\leftscriptspace{e}{}{\Gamma}\left(k_t\right)=0$ the incident field in \eqref{equ:line_source_PEC_incident_reflected_Ey}-\eqref{equ:line_source_PEC_incident_reflected_Hz} is reduced to the field produced by a line source and its image, positioned symmetrically $d+x'$ below the PEC \textcolor{black}{(Fig. \ref{fig:PhysicalConfigurationThreeSources}(c))}.

As in Subsubsection \ref{subsubsec:results:line_source:design}, we compare \eqref{equ:TE_spectral_domain_Ey}-\eqref{equ:TE_spectral_domain_Hz} with \eqref{equ:line_source_PEC_incident_reflected_Ey}-\eqref{equ:line_source_PEC_incident_reflected_Hz} in the \emph{source-free region} $x'<x<0$ (i.e., where $x_<=x'$ and $x_>=x$) to extract the source-related spectrum. This results in
\begin{eqnarray}
\leftscriptspace{e}{}{f}\left(k_t\right)=\dfrac{e^{-j\beta\left(d+x'\right)}-e^{j\beta\left(d+x'\right)}}
		{e^{j\beta d}-\leftscriptspace{e}{}{\Gamma}\left(k_t\right)e^{-j\beta d}}.
\label{equ:line_source_PEC_source_spectrum}
\end{eqnarray}

A careful examination of the derivation in Section \ref{sec:theory} reveals that the dependency of the source-related spectrum in the reflection coefficient does not affect the conditions for local impedance \textcolor{black}{equalization} (Table \ref{tab:design_rules} / Step 1), as in the source-free region, this dependency is the same for all fields (e.g., \eqref{equ:line_source_PEC_incident_reflected_Ey}-\eqref{equ:line_source_PEC_incident_reflected_Hz}). This means that enforcing local impedance \textcolor{black}{equalization} \eqref{equ:TE_local_impedance_matching_explicit_2} on the fields \eqref{equ:line_source_PEC_incident_reflected_Ey}-\eqref{equ:line_source_PEC_incident_reflected_Hz} in that region will result in the same expression for the reflection coefficient \eqref{equ:TE_reflection_coefficient}. Utilizing this, the incident and reflected fields can be completely evaluated in region 1 ($d<x<0$), and the dimensionless total field at $x\rightarrow 0^-$ will thus be given by \eqref{equ:TE_spectral_domain_F_minus}, reading
\begin{eqnarray}
\leftscript{e}{}{F}^{-}\left(z\right) \!=\! \frac{1}{2\pi}\!\!\int\limits_{-\infty}^{\infty}\!\!dk_t
	\dfrac{\sin\left[\beta\left(d+x'\right)\right]}
		{j\beta\cos\left(\beta d\right)-k\cos\theta_0\sin\left(\beta d\right)}e^{jk_tz}.
\label{equ:line_source_PEC_spectral_domain_F_minus}
\end{eqnarray}

As the poles of $\leftscriptspace{e}{}{f}\left(k_t\right)$ and the integrand of $\leftscript{e}{}{F}^{-}\left(z\right)$ coincide, the pole contributions to the integral \eqref{equ:line_source_PEC_spectral_domain_F_minus} indicate the guided and leaky modal fields in region 1 \cite{FelsenMarcuvitz1973, Holloway2012_1}. However, as these poles are complex ($\Im\left\lbrace k_{t,\mathrm{pole}}\right\rbrace\neq0$), their presence do not, in general, introduce significant difficulties to numerical evaluation of the integral. Hence, Step 1 of Table \ref{tab:design_rules} may be completed by evaluating $\left|\leftscript{e}{}{F}^{-}\left(z\right)\right|$ and $\varphi_{-}\left(z\right)$ from \eqref{equ:line_source_PEC_spectral_domain_F_minus}.

As in Subsubsection \ref{subsubsec:results:line_source:design} we follow \eqref{equ:TE_window_function_explicit} [Table \ref{tab:design_rules} / Step 2] and \eqref{equ:TE_phase_variation_plus} [Table \ref{tab:design_rules} / Step 3] to establish the shape of the virtual aperture and the phase compensation of the HMS, respectively, leading to the formulation of the HMS surface impedance and surface admittance
\begin{eqnarray}
\left\lbrace
\begin{array}{l}
\vspace{5pt}
\!\!\!\!\leftscript{e}{}{Z}_{se}\left(z\right)=-j\dfrac{\leftscript{e}{}{Z}_0}{2}
	\cot\left[\dfrac{kz\left(\sin\theta_0-\sin\theta_i\left(z\right)\right)+\xi_0}{2}\right]
	 \\
\!\!\!\!\leftscriptspace{e}{}{Y}_{sm}\left(z\right)=-j\dfrac{\leftscriptspace{e}{}{Y}_0}{2}
	\cot\left[\dfrac{kz\left(\sin\theta_0-\sin\theta_i\left(z\right)\right)+\xi_0}{2}\right],
\end{array}\!\!\!\!\!
\right.
\label{equ:line_source_PEC_impedance_admittance_final}
\end{eqnarray}
where we have used again the definition of the equivalent angle of incidence \eqref{equ:equivalent_angle_incidence}. Executing these steps enable the assessment of the transmitted fields as well.

\subsubsection{Virtual Aperture Engineering}
\label{subsubsec:results:line_source_PEC:aperture}
As the total field at $x\rightarrow 0^-$ is now an outcome of multiply-reflected field interference, it seems that controlling the shape of the virtual aperture window function becomes an even more difficult task. However, the introduction of the PEC reflector actually enhances our ability to control this function. As discussed in Subsubsection \ref{subsubsec:results:line_source:aperture}, in many cases the total field on the lower facet of the HMS can be approximated by the incident field. When the PEC is present, the incident field is created by an interference between the source and its image. Therefore, by controlling the relative position of the source and the PEC, we may affect this interference pattern, and subsequently the variation of the total field magnitude at $x\rightarrow 0^-$.

As an example, we may utilize the fact that in the absence of the HMS, the distance between the source and the PEC, $\left(d+x'\right)$, is related to the internal angle $\theta_{\mathrm{int}}$ in which constructive interference occurs in region 1 (Fig. \ref{fig:PhysicalConfigurationThreeSources}(c)) via \cite{Epstein2013_3}
\begin{eqnarray}
	\left(d+x'\right)\cos\theta_{\mathrm{int}} = \left(2n+1\right)\dfrac{\lambda}{4}, & n\in\mathbb{Z}.
\label{equ:line_source_PEC_constructive_interference}
\end{eqnarray}
Formation of two lobes travelling towards $\pm \theta_{\mathrm{int}}$ within region 1 should, in general, broaden the effective interaction length of the source field with the HMS on its lower facet, which, in turn, may enhance its aperture efficiency. As shall be demonstrated in the following Subsubsection, \eqref{equ:line_source_PEC_constructive_interference} can be used as an initial aperture engineering step, by which the suitable HMS-PEC distance $d$ is determined for given source position $x'$ and desirable constructive interference direction $\theta_{\mathrm{int}}$.

\subsubsection{Numerical and Semi-analytical Results}
\label{subsubsec:results:line_source_PEC:results}
\begin{figure}[!t]
\centering
\includegraphics[width=5.5cm]{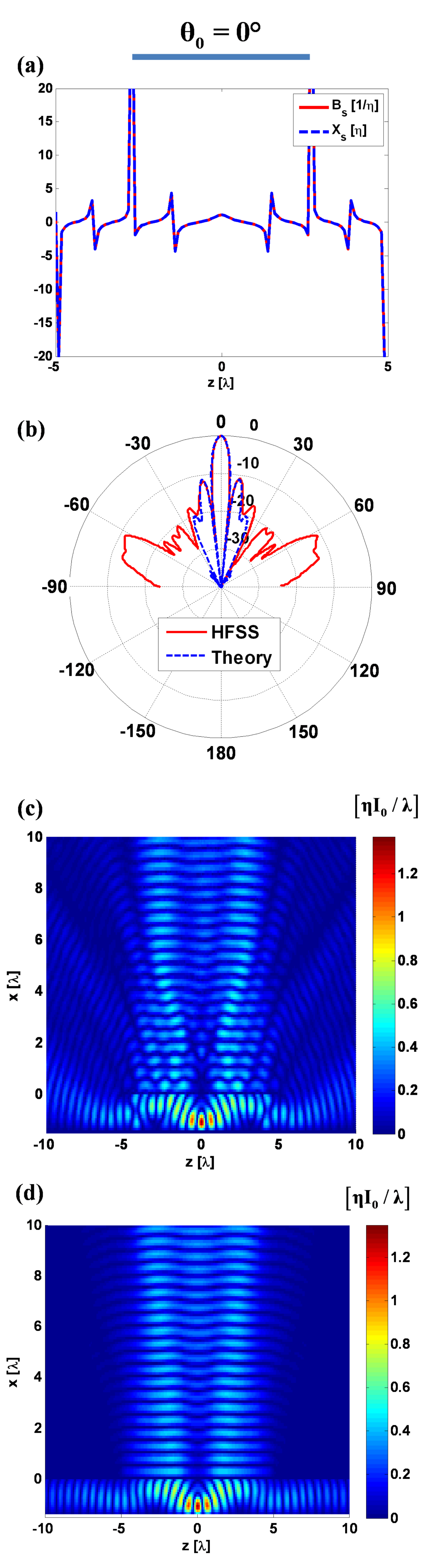}
\caption{Comparison between the theoretically predicted performance and the results of HFSS simulations of the HMS designed to convert the fields produced by an electric line source situated at $x'=-\lambda$ in front of a PEC at $x=-d=-1.5\lambda$ to directive radiation towards $\theta_0=0^{\circ}$. (a) Required surface reactance $X_s\left(z\right)$ (blue dashed line) and surface susceptance $B_s\left(z\right)$ (red solid line) calculated from \eqref{equ:line_source_PEC_impedance_admittance_final}. (b) Theoretically predicted (blue dashed line) and HFSS-simulated (red solid lines) normalized radiation pattern (dB scale). (c) Real part of the electric field phasor $\left|\Re\left\lbrace E_y\left(x,z\right)\right\rbrace\right|$ as simulated by HFSS. (d) Theoretical (semi-analytical) prediction of $\left|\Re\left\lbrace E_y\left(x,z\right)\right\rbrace\right|$.}
\label{fig:LineSourcePECResults}
\end{figure}

To verify the design procedure for the configuration considered in this Subsection, we have designed and simulated a Huygens metasurface according to the prescribed procedure (with $\xi_0=0$), designated to convert the fields produced by an electric line source at ${x=x'=-\lambda}$, positioned in front of a PEC at $x=-d$, to directive radiation towards $\theta_0=0^{\circ}$ (Fig. \ref{fig:LineSourcePECResults}). 

Following the discussion in Subsubsection \ref{subsubsec:results:line_source_PEC:aperture}, we aimed at harnessing the PEC reflector to improve the theoretical limit for aperture efficiency for a line source $\left|x'\right|=\lambda$ below the HMS, calculated in Subsubsection \ref{subsubsec:results:line_source:results} to be $71\%$ for $\theta_0=0^{\circ}$. To that end we positioned the PEC such that constructive interference would take place at $\theta_{\mathrm{int}}=60^{\circ}$ in region 1, yielding, utilizing \eqref{equ:line_source_PEC_constructive_interference} with $n=0$, an HMS-PEC distance of $d=1.5\lambda$. Although other values of $\theta_{\mathrm{int}}$ also broaden the effective interaction length of the incident field and the HMS, we have found using the semi-analytically estimated fields (Appendix \ref{app:asymptotic_evaluation}) that $\theta_{\mathrm{int}}=60^{\circ}$ yields an optimal result for the $10\lambda$-long HMS under consideration.

Fig. \ref{fig:LineSourcePECResults}(a) presents the surface reactance and surface susceptance required to implement the desirable HMS. As in Subsubsection \ref{subsubsec:results:line_source:results}, the design procedure output was used to implement the HMS in ANSYS HFSS (following the same procedure and unit cell structure, \textit{cf.} Appendix \ref{app:HFSS_implementation}), as well as to assess the predicted reflected and transmitted fields (following the semi-analytical approach of Appendix \ref{app:asymptotic_evaluation}).

Fig. \ref{fig:LineSourcePECResults}(b) compares between the normalized radiation patterns evaluated using the HFSS simulation (red solid line) and the semi-analytical theoeretical calculations (blue dashed line), showing a very good agreement between the theory and simulation for the angular range $\theta\in\left(-30^{\circ},30^{\circ}\right)$. As implied by the simulated (Fig. \ref{fig:LineSourcePECResults}(c)) and theoretical (Fig. \ref{fig:LineSourcePECResults}(d)) field plots, most of the discrepancy at large angles ($\theta>60^{\circ}$) originate from the fact that the semi-analytical approximations neglect the contribution of fields incident at the plane $x=0$ outside the metasurface $\left|z\right|>L/2$ to the radiation in region 2 (Appendix \ref{app:asymptotic_evaluation}).

Fig. \ref{fig:LineSourcePECResults}(c)-(d) highlight two additional properties of the line-source/PEC configuration. First, these two subfigures indicate in a clear manner that the profile of the virtual aperture $x\rightarrow 0^+$ indeed follows the total field magnitude at $x\rightarrow 0^-$. The two spots formed on the lower facet of the HMS due to the interference between the line source and its image (Subsubsection \ref{subsubsec:results:line_source_PEC:aperture}) are clearly translated into two dominant beams originating from the same positions on the upper facet. Second, as discussed briefly in Subsubsection \ref{subsubsec:results:line_source_PEC:design}, the expression for the total field in region 1 \eqref{equ:line_source_PEC_spectral_domain_F_minus} contains complex poles, and indeed the field plots indicate that the examined structure supports leaky modes. This is particularly pronounced in the theoretical prediction Fig. \ref{fig:LineSourcePECResults}(d), as the semi-analytical formulation used assumes the HMS is of infinite extent over the $z$ axis when evaluating the fields in region 1 (Appendix \ref{app:asymptotic_evaluation}), thus facilitating the observed long-range guidance of moderately leaking modes.

\begin{table}[!t]
\renewcommand{\arraystretch}{1.3}
\caption{Performance Parameters of the HMS Directive Radiator Excited by a Line Source in front of a PEC, Corresponding to Fig. \ref{fig:LineSourcePECResults}}
\label{tab:line_source_PEC_HMS_performance}
\centering
\begin{tabular}{l|c c c}
\hline \hline
& HFSS & Theory & Relative Performance \\
\hline \hline \\[-1.3em]	 
	Half-Power Beam Width
	 & $5.4^{\circ}$ & $5.4^{\circ}$ & $100\%$
	 	 	\\	\hline 
	Aperture Efficiency
	 & $93\%$ & $93\%$ & $100\%$
	 	 	 \\ \hline 
	Peak Directivity
	 & $38.9$ & $54.5$ & $71\%$
	 	 	 	 \\		 	
\hline \hline
\end{tabular}
\end{table}

Table \ref{tab:line_source_PEC_HMS_performance} summarizes the performance of the ELS/PEC excited HMS as a directive radiator. In consistency with Fig. \ref{fig:LineSourcePECResults}(b), the theoretically predicted and numerically simulated HPBW perfectly agree. As anticipated in Subsubsection \ref{subsubsec:results:line_source_PEC:aperture}, the aperture efficiency has indeed increased from $71\%$ for the ELS excited HMS (Table \ref{tab:line_source_HMS_performance}) to $93\%$ after introduction of the PEC, due to the utilization of the interference between the source and its image to broaden the incident field profile. An enhanced directivity is also recorded, where we note that when the PEC is present no power is lost due to radiation to region 1. It should be also noted that the theoretically predicted directivity is clearly overestimated, as is does not account for the fraction of the source power reaching region 2 without interacting with the HMS (i.e., via $\left|z\right|>L/2$, \textit{cf.} Appendix \ref{app:asymptotic_evaluation}), which has non-negligible contribution, as the numerical simulations reveal (Fig. \ref{fig:LineSourcePECResults}(c)).

\section{Conclusion}
\label{sec:conclusion}
We have presented a \textcolor{black}{detailed} formulation of a design procedure for scalar Huygens metasurface directive radiators, applicable for an arbitrary 2D source excitation. Our derivation reveals that satisfaction of two physical conditions is sufficient to guarantee that the designed HMS is passive and lossless: local power conservation and local impedance \textcolor{black}{equalization}. By expressing the incident, reflected and transmitted fields via their plane-wave spectrum, and utilizing the slowly-varying \textcolor{black}{envelope} approximation, we have shown that enforcing local impedance \textcolor{black}{equalization} results in a Fresnel-like reflection for the various spectral components. Furthermore, enforcing local power conservation dictates that the profile of the virtual aperture forming the transmitted directive radiation follows the magnitude of the total (incident and reflected) excitation fields. At the end of the design procedure, the fields at all regions may be assessed semi-analytically, and the required variation of the surface impedance and surface admittance along the HMS is prescribed.

We have verified our formulation using three different source configurations, showing good agreement between the semi-analytical predictions and finite-element simulation results executed by implementing the HMS as consecutive unit cells in ANSYS HFSS, yet to be optimized in a dedicated future work. Means to control the virtual aperture via modification of the source excitation were discussed and demonstrated as well.

The proposed design procedure establishes a flexible and robust foundation for the design and verification of novel antennas, allowing exploration of a vast variety of excitation forms via a single uniform formalism. Moreover, the derivation provides insight regarding the physical requirements to achieve passive lossless Huygens metasurfaces with desirable functionality, shedding light on previously investigated HMSs as well as indicating possible directions for future HMS applications. 


%

\appendices
\section{Derivation of HMS Surface Reactance and Susceptance for TM-Polarized Incident Fields}
\label{app:theory_TM_excitation}
For completeness, we provide here 
the final results of the derivation of the surface reactance required to implement the dual HMS, converting an arbitrary TM-polarized source field to directive radiation towards $\theta_0$ (derivable by duality \cite{Chew1990}). 

As denoted in Subsection \ref{subsec:theory:formulation}, when the excitation field is TM-polarized, the nonvanishing field components are $H_y\left(x,z\right)$, $E_z\left(x,z\right)$, and $E_x\left(x,z\right)$; the scalar surface impedance only induces \emph{electric} currents in the $z$ direction; and the scalar surface admittance only induces \emph{magnetic} currents in the $y$ direction. Analogously to \eqref{equ:TE_spectral_domain_Ey}-\eqref{equ:TE_spectral_domain_Hz}, the tangential magnetic fields in the \emph{source-free region} $x>x'$ can be generally formulated in the spectral domain as
\begin{eqnarray}
\left\lbrace
\begin{array}{l}
\!\!\!\!H_y^{\mathrm{inc}}\left(x,z\right) = -k \dfrac{I_0}{2\pi}\displaystyle\int\limits_{-\infty}^{\infty}\dfrac{dk_t}{2\beta}
	\leftscript{m}{}{f}\left(k_t\right)e^{-j\beta x}e^{jk_tz} \\
\!\!\!\!H_y^{\mathrm{ref}}\left(x,z\right) = k \dfrac{I_0}{2\pi}\displaystyle\int\limits_{-\infty}^{\infty}\dfrac{dk_t}{2\beta}
	\leftscriptspace{m}{}{\Gamma}\left(k_t\right)\leftscript{m}{}{f}\left(k_t\right)e^{j\beta x}e^{jk_tz} \\
\!\!\!\!H_y^{\mathrm{trans}}\left(x,z\right) = -k \dfrac{I_0}{2\pi}\displaystyle\int\limits_{-\infty}^{\infty}\dfrac{dk_t}{2\beta}
	\leftscriptspace{m}{}{T}\left(k_t\right)e^{-j\beta x}e^{jk_tz},
\end{array}\!\!\!\!
\right.
\label{equ:TM_spectral_domain_Hy}
\end{eqnarray}
and the respective tangential electric fields can be derived via \eqref{equ:TM_Maxwell_equations};
the $m$ left superscript denotes TM-HMS related quantities. 
The scalar surface impedance and admittance may be written in analogy to \eqref{equ:TE_impedance_admittance_discontinuity} as 
\begin{eqnarray}
\left\lbrace
\begin{array}{l}
\vspace{5pt}
\!\!\!\!\leftscript{m}{}{Z}_{se}\left(z\right)\!=\!\dfrac{1}{2}
\dfrac{E_z^{\mathrm{trans}}\left(0,z\right)+\left[E_z^{\mathrm{inc}}\left(0,z\right)+E_z^{\mathrm{ref}}\left(0,z\right)\right]}{H_y^{\mathrm{trans}}\left(0,z\right)-\left[H_y^{\mathrm{inc}}\left(0,z\right)+H_y^{\mathrm{ref}}\left(0,z\right)\right]} \\
\!\!\!\!\!\leftscriptspace{m}{}{Y}_{sm}\left(z\right)\!=\!\dfrac{1}{2}
\dfrac{H_y^{\mathrm{trans}}\left(0,z\right)+\left[H_y^{\mathrm{inc}}\left(0,z\right)+H_y^{\mathrm{ref}}\left(0,z\right)\right]}{E_z^{\mathrm{trans}}\left(0,z\right)-\left[E_z^{\mathrm{inc}}\left(0,z\right)+E_z^{\mathrm{ref}}\left(0,z\right)\right]}.
\end{array}\!\!\!\!\!\!\!
\right.
\label{equ:TM_impedance_admittance_discontinuity}
\end{eqnarray}

The transmitted tangential fields are defined using a virtual aperture window function $\leftscriptspace{m}{}{W\left(x,z\right)}$, similarly to \eqref{equ:TE_window_function_Ey},
\begin{eqnarray}
H_y^{\mathrm{trans}}\left(x,z\right) = -k I_0 \leftscriptspace{m}{}{W\left(x,z\right)}e^{-jkx\cos\theta_0}e^{-jkz\sin\theta_0},
\label{equ:TM_window_function_Hy}
\end{eqnarray}
which we assume to satisfy the slowly-varying \textcolor{black}{envelope} condition \eqref{equ:TE_slowly_varying_W} as $x\rightarrow 0^{+}$ (with $e$ superscript replaced by $m$).

The dimensionless field quantities are defined for TM-polarized source field as
\begin{eqnarray}
\left\lbrace
\begin{array}{l}
\vspace{5pt}
\leftscript{m}{}{F}_{\!\!E}^{-}\left(z\right) \triangleq \dfrac{1}{I_0k\eta\cos\theta_0}
	\left[E_z^{\mathrm{inc}}\left(0,z\right)+E_z^{\mathrm{ref}}\left(0,z\right)\right] \\
\vspace{5pt}
\leftscript{m}{}{F}_{\!\!H}^{-}\left(z\right) \triangleq -\dfrac{1}{I_0k}
	\left[H_y^{\mathrm{inc}}\left(0,z\right)+H_y^{\mathrm{ref}}\left(0,z\right)\right] \\
\leftscript{m}{}{F}^{+}\left(z\right) \triangleq \leftscriptspace{m}{}{W\left(0,z\right)}e^{-jkz\sin\theta_0}
\end{array}\!\!\!\!\!\!
\right.
\label{equ:TM_dimensionless_quantities}
\end{eqnarray}
using which local impedance \textcolor{black}{equalization} and local power conservation conditions retain the same formulae as in \eqref{equ:TE_local_impedance_matching} and \eqref{equ:TE_local_power_conservation_impedance_matched}, with the $e$ superscript replaced by $m$, likewise defining the analogue of \eqref{equ:TE_dimensionless_quantities_passive_HMS}.

Finally, the surface reactance for the TM case is given by
\begin{eqnarray}
\left\lbrace
\begin{array}{l}
\vspace{5pt}
\!\!\!\!\leftscript{m}{}{Z}_{se}\left(z\right)=-j\dfrac{\leftscript{m}{}{Z}_0}{2}
	\cot\left[\dfrac{\varphi_{-}\left(z\right)-\varphi_{+}\left(z\right)}{2}\right]
	 \\
\!\!\!\!\leftscriptspace{m}{}{Y}_{sm}\left(z\right)=-j\dfrac{\leftscriptspace{m}{}{Y}_0}{2}
	\cot\left[\dfrac{\varphi_{-}\left(z\right)-\varphi_{+}\left(z\right)}{2}\right],
\end{array}\!\!\!\!\!
\right.
\label{equ:TM_impedance_admittance_final}
\end{eqnarray}
where $\leftscript{m}{}{Z}_0=1/\leftscriptspace{m}{}{Y}_0=\eta\cos\theta_0$ is the wave impedance of a TM-polarized plane-wave propagating in region 2 at an angle of $\theta_0$ with respect to the $x$ axis (See \eqref{equ:transmitted_wave_impedance}). Despite the different wave impedance for TE and TM polarizations, the reflection coefficient arising from local impedance \textcolor{black}{equalization} remains the same. Consequently, explicit evaluation of the scattered fields and the phases $\varphi_{\pm}\left(z\right)$ may be obtained by using \eqref{equ:TE_reflection_coefficient}-\eqref{equ:TE_impedance_admittance_final_explicit} with the $e$ left superscript replaced by $m$.

%
%

\section{Implementation and Simulation of the HMSs in ANSYS HFSS}
\label{app:HFSS_implementation}
The Huygens metasurface designs investigated in Subsubsections \ref{subsubsec:results:line_source:results} and \ref{subsubsec:results:line_source_PEC:results} were simulated using a commerically available finite-element solver (ANSYS HFSS) to verify the theoretical predictions (calculated using the continuous equivalent surface impedance and admittance) via comparison to a more realistic implementation of the HMS.

\begin{figure}[!t]
\centering
\includegraphics[width=6.5cm]{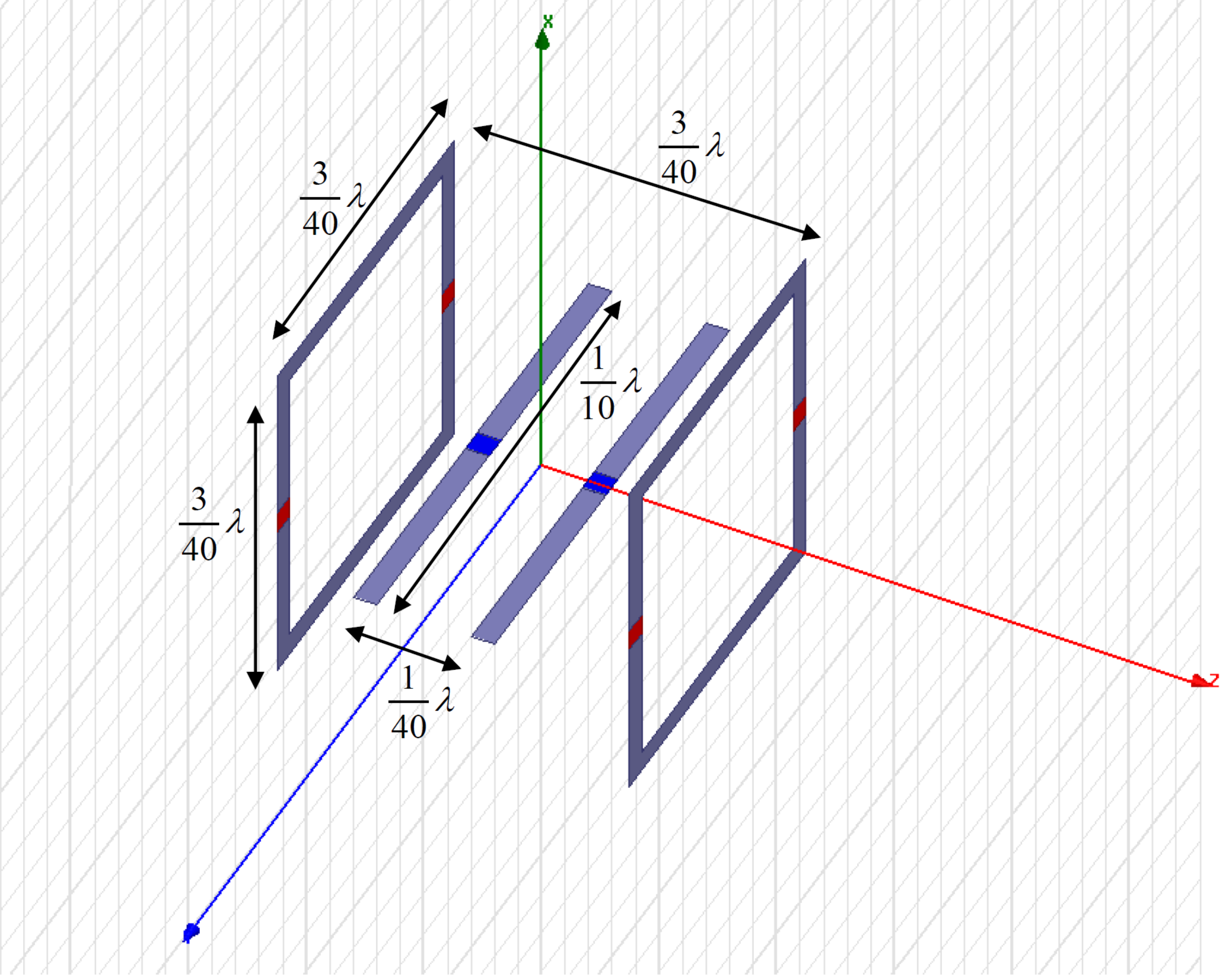}
\caption{Physical configuration of a symmetrical unit cell for HMS implementation. The unit cell is comprised of two identical PEC squared loops loaded by lumped capacitors (marked in red), and two identical PEC wires loaded with either lumped capacitors or lumped inductors (marked in blue). The magnetic dipole formed by the loaded loops is oriented parallel to the $z$-axis while the electric dipole formed by the loaded wires is oriented parallel to the $y$-axis.}
\label{fig:UnitCell}
\end{figure}

The simulated $10\lambda$-long HMSs were implemented using one-hundred $\lambda/10$-long unit cells, where the frequency of the time-harmonic excitations was $f=1.5\mathrm{GHz}$. Fig. \ref{fig:UnitCell} presents the physical configuration of a unit cell, along with its dimensions. Each unit cell is composed of two square PEC loops, loaded by identical lumped capacitors, and two PEC wires, loaded either by identical lumped capacitors or by identical lumped inductors, positioned symmetrically with respect to the center of the cell. The magnetic dipole formed 
by the loaded loops is oriented parallel to the $z$-axis while the electric dipole formed by the loaded wires is oriented parallel to the $y$-axis, indicating that the unit cell is only sensitive to TE-polarized fields (See Subsubsection \ref{subsubsec:theory:formulation:spectral_decomposition}). We simulate a 2D environment by placing two PEC surfaces at $y=\pm \lambda/20$.

The surface impedance and admittance of a unit cell for a given lumped loading is evaluated by simulating the response of an infinite periodic array of identical unit cells to a normally incident (TE-polarized) plane-wave excitation. Utilizing the impedance matrix calculated by HFSS for the scattering problem, and the circuit model introduced in \cite{Selvanayagam2013_1}, we extract the equivalent surface impedance and admittance of the cell. Variation of the lumped loading of the wires and loops induces, respectively, variation of the equivalent surface impedance and admittance of the cell \cite{Tretyakov2003}. This facilitates the generation of a lookup table, matching pairs of surface reactance and susceptance to sets of lumped capacitors or inductors. Invoking the principle of local periodicity, a required polarizability variation such as the ones in Fig. \ref{fig:LineSourceResults}(a)-(c) and Fig. \ref{fig:LineSourcePECResults}(a) can be sampled at a suitable unit-cell-length period ($\lambda/10$ in our case) and translated using the lookup table into a set of unit cells with prescribed lumped loadings \cite{Selvanayagam2013,Pfeiffer2013,Pfeiffer2013_1}.  

Although the electric and magnetic dipoles are orthogonal, and ideally should exhibit decoupled responses to orthogonal electric and magnetic fields, we discovered that due to the finite size of the structures, some coupling does exist. This is especially pronounced when the lumped loading is such that the element is near resonance at the working frequency $f$. 
Thus, to enhance the accuracy of the lookup table, we have chosen to simulate the unit cell as a whole 
 and to relate pairs of reactance/susceptance to pairs of lumped elements.  

\section{Semi-analytical Prediction of HMS Performance}
\label{app:asymptotic_evaluation}
To establish an effective engineering methodology based on the design procedure introduced in Section \ref{sec:theory}, it is desirable to develop computationally efficient tools to estimate the HMS performance before resorting to numerical simulation tools for final optimization. In Subsubsection \ref{subsubsec:results:line_source:results} and Subsubsection \ref{subsubsec:results:line_source_PEC:results} we have employed such tools, in the form of semi-analytical formulae, to assess the agreement between simulated results of detailed design and the predictions of the more idealized theory. In this Appendix we describe the details regarding the evaluation of these semi-analytical formulae, as well as the key approximations used.

Formally, the theoretical derivation of Section \ref{sec:theory} is valid only for infinitely long HMSs. If the HMS is finite, the problem is no longer uniform (the configuration is not separable), and the spectral analysis employed herein cannot be carried out. Nevertheless, in practice, if an efficient conversion of the source power to directive radiation is to be achieved, the HMS length $L$ should be judiciously chosen such that most of the interaction between the source power and an infinitely long HMS takes place over a finite region $\left|z\right|<L/2$ of the $yz$ plane. Assuming this is indeed the case (\emph{Assumption 1}), we may treat the HMS as infinite over the $z$ axis without introducing significant errors with respect to the realistic finite-length implementation. This, in turn, facilitates the utilization of \eqref{equ:TE_spectral_domain_F_minus} to evaluate the fields in region 1, disregarding any effects arising from the edges of the HMS, e.g. diffraction or a discontinuity in the reflection coefficient.   

A second assumption essential for the validity of our derivation is that the magnitude of the total field impinging upon the HMS at $x\rightarrow 0^-$, including the reflection induced by local impedance \textcolor{black}{equalization}, varies moderately such that the slowly-varying \textcolor{black}{envelope} condition \eqref{equ:TE_slowly_varying_W} is satisfied. Utilizing this assumption (\emph{Assumption 2}), we may use \eqref{equ:TE_window_function_explicit}-\eqref{equ:TE_transmission_spectral_domain} to evaluate the fields on the upper facet of the HMS ($x\rightarrow 0^+$).
However, when evaluating the fields in region 2, we would like to account for the fact that the HMS is of finite length $L$, which may have significant effects on the radiated fields. Hence, we truncate the virtual aperture window function $W\left(0,z\right)$ at the edges of the implemented HMS; practically, the integral of \eqref{equ:TE_transmission_spectral_domain} is executed with the infinite limits replaced by $\pm L/2$. 
This is equivalent to assuming that the fields at the HMS plane $x=0$ vanish in the region where the HMS is absent $\left|z\right|>L/2$. \emph{Assumption 1} above facilitates this approximation.

These assumptions are used to evaluate the spectral content of the reflected and transmitted fields, and consequently, via the spectral integrals \eqref{equ:TE_spectral_domain_Ey}-\eqref{equ:TE_spectral_domain_Hz}, the fields everywhere above and below the HMS (Fig. \ref{fig:LineSourceResults}(j)-(l) and Fig. \ref{fig:LineSourceResults}(d)).

Nonetheless, when far-field evaluation is required, such as for the assessment of the radiation pattern, straightforward spectral integration is problematic due to the highly oscillating nature of the radiation integrals as $x,z\rightarrow \pm\infty$. Hence, for the evaluation of far-field radiated power we apply the steepest-descent-path method, yielding closed-form asymptotic approximations for the spectral integrals. For the typical spectral integrals \eqref{equ:TE_spectral_domain_Ey}-\eqref{equ:TE_spectral_domain_Hz}
\begin{eqnarray}
I_{\pm}\left(x,z\right) = \displaystyle\int\limits_{-\infty}^{\infty}\dfrac{dk_t}{2\beta}
	g\left(k_t\right)e^{\pm j\beta x}e^{jk_tz}
\label{equ:field_spectral_integral}
\end{eqnarray}
the asymptotic evaluation at a point $\left(x=r\cos\theta,z=r\sin\theta\right)$ where $kr\rightarrow\infty$ and $\theta\in\left(-\pi,\pi\right]$ is given by \cite{FelsenMarcuvitz1973,Epstein2013_3}
\begin{eqnarray}
I_{\pm}\left(r,\theta\right) \sim \sqrt{\dfrac{\pi}{2kr}}g\left(k_t=k\sin\alpha_{\pm}\right)e^{-jkr}e^{j\pi/4}
\label{equ:field_spectral_integral_SDP}
\end{eqnarray}
where $\alpha_{-}=-\theta$ for $x>0$ and $\alpha_{+}=\theta+\pi$ for $x<0$, assuming the sign attached to $\beta$ in the exponent of \eqref{equ:field_spectral_integral} satisfies the radiation condition at the respective regions.

\section*{Acknowledgment}
A.E. gratefully acknowledges the support of The Lyon Sachs Postodoctoral Fellowship Foundation as well as The Andrew and Erna Finci Viterbi Fellowship Foundation of the Technion - Israel Institute of Technology, Haifa, Israel.

\ifCLASSOPTIONcaptionsoff
  \newpage
\fi

\begin{IEEEbiography}[{\includegraphics[width=1in,height=1.25in,clip,keepaspectratio]{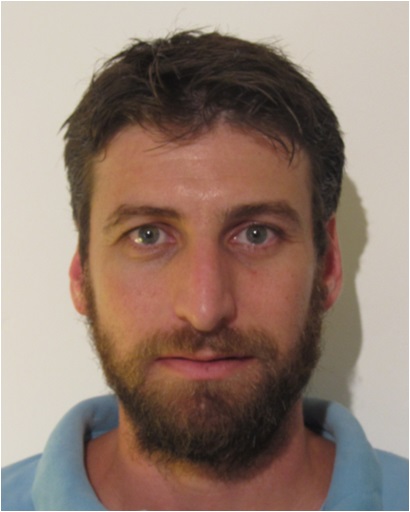}}]{Ariel Epstein}
(S'12-M'14) received the B.A. degree in computer science from the Open University of Israel, Raanana, the B.A. degree in physics, and the B.Sc. and Ph.D. degrees in electrical engineering from the Technion - Israel Institute of Technology, Haifa, Israel, in 2000, 2003, and 2013 respectively. 
He is currently a Lyon Sachs Postdoctoral Fellow at the Department of Electrical and Computer Engineering in the University of Toronto, Ontario, Canada. 

His current research interests include utilization of electromagnetic theory, with emphasis on analytical techniques, for novel applications, e.g. electromagnetic metamaterials for antennas and microwave components.

Dr. Epstein was the recipient of the Young Scientist Best Paper Award in the URSI Commission B International Symposium on Electromagnetic Theory (EMTS2013), held in Hiroshima, Japan on May 2013, as well as the Best Poster Award at the 11th International Symposium on Functional $\pi$-electron Systems (F$\pi$-11), held in Arcachon, France on June 2013.
\end{IEEEbiography}

\begin{IEEEbiography}[{\includegraphics[width=1in,height=1.25in,clip,keepaspectratio]{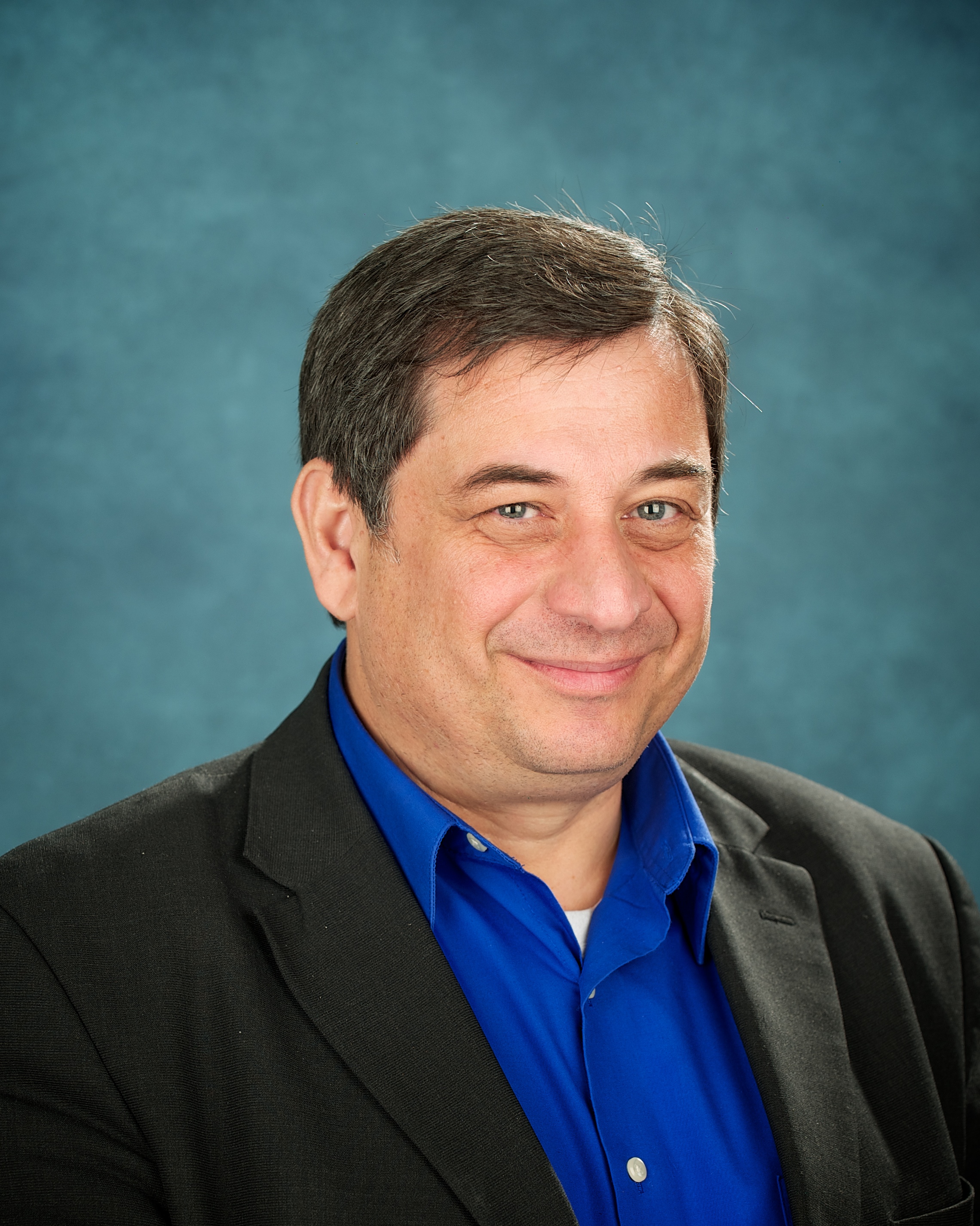}}]{George V. Eleftheriades}
(S'86-M'88-SM'02-F'06) earned the M.S.E.E. and Ph.D. degrees in electrical engineering from the University of Michigan, Ann Arbor, MI, USA, in 1989 and 1993, respectively, where he was involved in the development of sub-millimeter-wave and THz technology sponsored by NASA. From 1994 to 1997, he was with the Swiss Federal Institute of Technology, Lausanne, Switzerland, where he was engaged in the design of millimeter and sub-millimeter-wave receivers for the European Space Agency. Currently, he is a Professor in the Department of Electrical and Computer Engineering at the University of Toronto, ON, Canada, where he holds the Canada Research Chair in Nano- and Micro-Structured Electromagnetic Materials and Applications and the Velma M. Rogers Graham Chair in Engineering.
He is a recognized international authority and pioneer in the area of negative-refractive-index metamaterials. These are man-made materials which have electromagnetic properties not found in nature. He introduced a method for synthesizing metamaterials using loaded transmission lines.
Together with his graduate students, he provided the first experimental evidence of imaging beyond the diffraction limit and pioneered several novel microwave components and antennas using these transmission-line based metamaterials. His research has impacted the field by demonstrating the unique electromagnetic properties of metamaterials; used in lenses, antennas, and other microwave components to drive innovation in fields such as defence, medical imaging, microscopy, automotive radar, and wireless telecommunications. Presently, he is leading a group  of graduate students and researchers in the areas of electromagnetic negative-refraction metamaterials, transformation optics, metasurfaces, small antennas and components for broadband wireless communications, novel antenna beam-steering techniques, plasmonic and nanoscale optical components, and fundamental electromagnetic theory.

Prof. Eleftheriades served as an Associate Editor for the IEEE TRANSACTIONS ON ANTENNAS AND PROPAGATION (AP). He also served as a member of the IEEE AP-Society administrative committee (AdCom) from 2007 to 2012 and was an IEEE AP-S Distinguished Lecturer from 2004 to 2009. He served as the General Chair of the 2010 IEEE International Symposium on Antennas and Propagation held in Toronto, ON, Canada. Papers that he coauthored have received numerous awards such as the 2009 Best Paper Award from the IEEE MICROWAVE AND WIRELESS PROPAGATION LETTERS,  twice the R. W. P. King Best Paper Award from the IEEE TRANSACTIONS ON ANTENNAS AND PROPAGATION (2008 and 2012), and the 2014 Best Paper Award from the IEEE ANTENNAS AND WIRELESS PROPAGATION LETTERS. His work has been cited more than 10,000 times, and his h-index is 47. He received the Ontario Premier’s Research Excellence Award and the University of Toronto's Gordon Slemon Award, both in 2001. In
2004 he received an E.W.R. Steacie Fellowship from the Natural Sciences and Engineering Research Council of Canada. He received the 2008 IEEE Kiyo Tomiyasu Technical Field Award “for pioneering contributions to the science and technological applications of negative-refraction electromagnetic materials.” In 2009, he was elected a Fellow of the Royal Society of Canada.
\end{IEEEbiography}



\vfill



\begin{thebibliography}{10}
\providecommand{\url}[1]{#1}
\csname url@samestyle\endcsname
\providecommand{\newblock}{\relax}
\providecommand{\bibinfo}[2]{#2}
\providecommand{\BIBentrySTDinterwordspacing}{\spaceskip=0pt\relax}
\providecommand{\BIBentryALTinterwordstretchfactor}{4}
\providecommand{\BIBentryALTinterwordspacing}{\spaceskip=\fontdimen2\font plus
\BIBentryALTinterwordstretchfactor\fontdimen3\font minus
  \fontdimen4\font\relax}
\providecommand{\BIBforeignlanguage}[2]{{%
\expandafter\ifx\csname l@#1\endcsname\relax
\typeout{** WARNING: IEEEtran.bst: No hyphenation pattern has been}%
\typeout{** loaded for the language `#1'. Using the pattern for}%
\typeout{** the default language instead.}%
\else
\language=\csname l@#1\endcsname
\fi
#2}}
\providecommand{\BIBdecl}{\relax}
\BIBdecl

\bibitem{Trentini1956}
G.~V. Trentini, ``Partially reflecting sheet arrays,'' \emph{IRE Trans.
  Antennas Propag.}, vol.~4, no.~4, pp. 666--671, 1956.

\bibitem{Pozar1997}
D.~M. Pozar, S.~D. Targonski, and H.~D. Syrigos, ``Design of millimeter wave
  microstrip reflectarrays,'' \emph{IEEE Trans. Antennas Propag.}, vol.~45,
  no.~2, pp. 287--296, Feb. 1997.

\bibitem{Sievenpiper1999}
D.~Sievenpiper, L.~Zhang, R.~F.~J. Broas, N.~Alexopolous, and E.~Yablonovitch,
  ``High-impedance electromagnetic surfaces with a forbidden frequency band,''
  \emph{IEEE Trans. Microw. Theory Techn.}, vol.~47, no.~11, pp. 2059--2074,
  Nov. 1999.

\bibitem{Romeu2000}
J.~Romeu and Y.~Rahmat-Samii, ``{Fractal FSS: a novel dual-band frequency
  selective surface},'' \emph{IEEE Trans. Antennas Propag.}, vol.~48, no.~7,
  pp. 1097--1105, July 2000.

\bibitem{Sarabandi2007}
K.~Sarabandi and N.~Behdad, ``A frequency selective surface with miniaturized
  elements,'' \emph{IEEE Trans. Antennas Propag.}, vol.~55, no.~5, pp.
  1239--1245, May 2007.

\bibitem{Hum2014}
S.~Hum and J.~Perruisseau-Carrier, ``Reconfigurable reflectarrays and array
  lenses for dynamic antenna beam control: A review,'' \emph{IEEE Trans.
  Antennas Propag.}, vol.~62, no.~1, pp. 183--198, Jan. 2014.

\bibitem{Maci2011}
S.~Maci, G.~Minatti, M.~Casaletti, and M.~Bosiljevac, ``Metasurfing: Addressing
  waves on impenetrable metasurfaces,'' \emph{IEEE Antennas Wireless Propag.
  Lett.}, vol.~10, pp. 1499--1502, 2011.

\bibitem{Holloway2012}
C.~L. Holloway, E.~F. Kuester, J.~a. Gordon, J.~O'Hara, J.~Booth, and D.~R.
  Smith, ``{An overview of the theory and applications of metasurfaces: the
  two-dimensional equivalents of metamaterials},'' \emph{IEEE Antennas Propag.
  Mag.}, vol.~54, no.~2, pp. 10--35, Apr. 2012.

\bibitem{Kildishev2013}
A.~V. Kildishev, A.~Boltasseva, and V.~M. Shalaev, ``Planar photonics with
  metasurfaces,'' \emph{Science}, vol. 339, no. 6125, 2013.

\bibitem{Yu2014}
N.~Yu and F.~Capasso, ``{Flat optics with designer metasurfaces.}'' \emph{Nat.
  Mater.}, vol.~13, no.~2, pp. 139--50, 2014.

\bibitem{Kuester2003}
E.~Kuester, M.~Mohamed, M.~Piket-May, and C.~Holloway, ``{Averaged transition
  conditions for electromagnetic fields at a metafilm},'' \emph{IEEE Trans.
  Antennas Propag.}, vol.~51, no.~10, pp. 2641--2651, Oct. 2003.

\bibitem{Tretyakov2003}
S.~Tretyakov, \emph{Analytical Modeling in Applied Electromagnetics}.\hskip 1em
  plus 0.5em minus 0.4em\relax Artech House, 2003.

\bibitem{Zhao2011}
Y.~Zhao, N.~Engheta, and A.~Al\'{u}, ``Homogenization of plasmonic metasurfaces
  modeled as transmission-line loads,'' \emph{Metamaterials}, vol.~5, no.
  2ג€“3, pp. 90 -- 96, 2011.

\bibitem{Pendry2012}
J.~B. Pendry, A.~Aubry, D.~R. Smith, and S.~A. Maier, ``Transformation optics
  and subwavelength control of light,'' \emph{Science}, vol. 337, no. 6094, pp.
  549--552, 2012.

\bibitem{Eleftheriades2012}
G.~V. Eleftheriades and M.~Selvanayagam, ``Transforming electromagnetics using
  metamaterials,'' \emph{IEEE Microw. Mag.}, vol.~13, no.~2, pp. 26--38, Mar.
  2012.

\bibitem{Martini2014}
E.~Martini, G.~Sardi, and S.~Maci, ``Homogenization processes and retrieval of
  equivalent constitutive parameters for multisurface-metamaterials,''
  \emph{IEEE Trans. Antennas Propag.}, vol.~62, no.~4, pp. 2081--2092, Apr.
  2014.

\bibitem{Balanis1997}
C.~Balanis, \emph{Antenna Theory : Analysis and Design}.\hskip 1em plus 0.5em
  minus 0.4em\relax New York: Wiley, 1997.

\bibitem{Pfeiffer2013}
C.~Pfeiffer and A.~Grbic, ``{Metamaterial Huygensג€™ surfaces: tailoring wave
  fronts with reflectionless sheets},'' \emph{Phys. Rev. Lett.}, vol. 110,
  no.~19, p. 197401, May 2013.

\bibitem{Selvanayagam2013}
M.~Selvanayagam and G.~V. Eleftheriades, ``Discontinuous electromagnetic fields
  using orthogonal electric and magnetic currentsfor wavefront manipulation,''
  \emph{Opt. Express}, vol.~21, no.~12, pp. 14\,409--14\,429, June 2013.

\bibitem{Selvanayagam2012}
------, ``An active electromagnetic cloak using the equivalence principle,''
  \emph{IEEE Antennas Wireless Propag. Lett.}, vol.~11, pp. 1226--1229, 2012.

\bibitem{Selvanayagam2013_1}
------, ``Experimental demonstration of active electromagnetic cloaking,''
  \emph{Phys. Rev. X}, vol.~3, p. 041011, Nov. 2013.

\bibitem{Jin2010}
P.~Jin and R.~Ziolkowski, ``{Metamaterial-inspired, electrically small Huygens
  sources},'' \emph{IEEE Antennas Wireless Propag. Lett.}, vol.~9, pp.
  501--505, 2010.

\bibitem{Niemi2013}
T.~Niemi, A.~O. Karilainen, and S.~A. Tretyakov, ``{Synthesis of polarization
  transformers},'' \emph{IEEE Trans. Antennas Propag.}, vol.~61, no.~6, pp.
  3102--3111, 2013.

\bibitem{Rapoport2013}
Y.~G. Rapoport, S.~Tretyakov, and S.~Maslovski, ``{Nonlinear active Huygens
  metasurfaces for reflectionless phase conjugation of electromagnetic waves in
  electrically thin layers},'' \emph{J. Electromagnet. Wave.}, vol.~27, no.~11,
  pp. 1309--1328, 2013.

\bibitem{Selvanayagam2014}
M.~Selvanayagam and G.~V. Eleftheriades, ``{Polarization control using tensor
  Huygens metasurfaces},'' 2014, to be published.

\bibitem{Holloway2005}
C.~Holloway, M.~Mohamed, E.~Kuester, and A.~Dienstfrey, ``{Reflection and
  transmission properties of a metafilm: with an application to a controllable
  surface composed of resonant particles},'' \emph{IEEE Trans. Electromagn.
  Compat.}, vol.~47, no.~4, pp. 853--865, 2005.

\bibitem{Fong2010}
B.~Fong, J.~Colburn, J.~Ottusch, J.~Visher, and D.~Sievenpiper, ``{Scalar and
  tensor holographic artificial impedance surfaces},'' \emph{IEEE Trans.
  Antennas Propag.}, vol.~58, no.~10, pp. 3212--3221, 2010.

\bibitem{Pfeiffer2013_1}
C.~Pfeiffer and A.~Grbic, ``{Cascaded metasurfaces for complete phase and
  polarization control},'' \emph{Appl. Phys. Lett.}, vol. 102, no.~23, p.
  231116, 2013.

\bibitem{Holloway2012_1}
C.~L. Holloway, D.~C. Love, E.~F. Kuester, J.~A. Gordon, and D.~A. Hill, ``{Use
  of generalized sheet transition conditions to model guided waves on
  metasurfaces/metafilms},'' \emph{IEEE Trans. Antennas Propag.}, vol.~60,
  no.~11, pp. 5173--5186, 2012.

\bibitem{FelsenMarcuvitz1973}
L.~B. Felsen and N.~Marcuvitz, \emph{Radiation and Scattering of Waves},
  1st~ed.\hskip 1em plus 0.5em minus 0.4em\relax Englewood Cliffs, N.J.:
  Prentice-Hall, 1973.

\bibitem{Chew1990}
W.~C. Chew, \emph{{Waves and Fields in Inhomogeneous Media}}.\hskip 1em plus
  0.5em minus 0.4em\relax New York: Van Nostrand Reinhold, 1990.

\bibitem{Epstein2013_3}
\BIBentryALTinterwordspacing
A.~Epstein, ``Rigorous electromagnetic analysis of optical emission of organic
  light-emitting diodes,'' Ph.D. dissertation, Dept. Elect. Eng., Technion -
  Israel Insititute of Technology, Haifa, Israel, 2013. [Online]. Available:
  \url{http://arielepstein.webs.com/Research/thesis.pdf}
\BIBentrySTDinterwordspacing

\bibitem{Lovat2006}
G.~Lovat, P.~Burghignoli, F.~Capolino, D.~Jackson, and D.~Wilton, ``{Analysis
  of directive radiation from a line source in a metamaterial slab with low
  permittivity},'' \emph{IEEE Trans. Antennas Propag.}, vol.~54, no.~3, pp.
  1017--1030, Mar. 2006.

\end{thebibliography}
\end{document}